\pdfoutput=1
\let\oldvec\vec% Store \vec in \oldvec
\documentclass{aa}
\let\vec\oldvec% Restore \vec from \oldvec

\newlength{\wfig}
\usepackage{graphicx}
\usepackage{amssymb,amsmath}
\usepackage{txfonts}
\usepackage{natbib}
\usepackage{rotating}
\usepackage{color, colortbl}
\usepackage{physics}
\renewcommand{\vec}[1]{\mathbf{#1}}

\definecolor{LightCyan}{rgb}{0.88,1,1}

\definecolor{Gray}{gray}{0.9}

\DeclareSymbolFont{mymathvariables}{OT1}{cmr}{m}{n}
\SetSymbolFont{mymathvariables}{normal}{OT1}{cmr}{m}{n}
\DeclareSymbolFontAlphabet{\mathnormal}{mymathvariables}
\DeclareMathSymbol{v}{\mathalpha}{mymathvariables}{118}

\begin{document}

\title{Solar type III radio burst time characteristics at LOFAR frequencies and the implications for electron beam transport}

   \author{Hamish A. S. Reid and Eduard P. Kontar}
   \institute{SUPA School of Physics and Astronomy,   University of Glasgow, G12 8QQ, United Kingdom}

\abstract
{Solar type III radio bursts contain a wealth of information about the dynamics of electron beams in the solar corona and the inner heliosphere; currently unobtainable through other means. However, the motion of different regions of an electron beam (front, middle and back) have never been systematically analysed before.}
{We characterise the type III burst frequency-time evolution using the enhanced resolution of LOFAR in the frequency range 30--70 MHz and use this to probe electron beam dynamics.}
{The rise, peak and decay times with a $\sim 0.2$ MHz spectral resolution were defined for a collection of 31 type III bursts.  The frequency evolution is used to ascertain the apparent velocities of the front, middle and back of the type III sources and the trends are interpreted using theoretical and numerical treatments.}
{The type III time profile was better approximated by an asymmetric Gaussian profile, not an exponential as previously used.  Rise and decay times increased with decreasing frequency and showed a strong correlation.  Durations were smaller than previously observed.  Drift rates from the rise times were faster than from the decay times, corresponding to inferred mean electron beam speeds for the front, middle and back of $0.2, 0.17, 0,15$~c, respectively.  Faster beam speeds correlate with smaller type III durations.  We also find type III frequency bandwidth decreases as frequency decreases.}
{The different speeds naturally explain the elongation of an electron beam in space as it propagates through the heliosphere.  The rate of expansion is proportional to the mean speed of the exciter; faster beams expand faster.  Beam speeds are attributed to varying ensembles of electron energies at the front, middle and back of the beam.}


\titlerunning{Type III frequency evolution}
\authorrunning{Reid \& Kontar}

\maketitle

\section{Introduction}

Type III radio bursts are understood to be signatures of accelerated electrons in the solar corona and interplanetary space.  Analysis of their frequency-time profile allows electron beam characteristics to be diagnosed close to the Sun where spacecraft are unable to make in situ plasma measurements.  The rapid change in the frequency of these radio bursts is attributed to the electron beams having high velocities, around $0.3c$, where $c$ is the speed of light. The time profiles at an individual frequency provides information about the spatial and energetic profiles of the electrons that participate in the plasma emission mechanism as well as the radio wave escape from the turbulent solar corona. 

The time profile of type III radio bursts has been studied by numerous authors \citep[e.g.][]{Hughes:1963aa,Aubier:1972aa,Evans:1973aa,Alvarez:1973aa,Barrow:1975aa,Poquerusse:1977aa,McLean:1985aa,Tsybko:1989aa,Melnik:2011aa} to investigate electron beam and background plasma parameters.  At a given frequency, it was suggested early on \citep[see][for a complete explanation]{Aubier:1972aa} that the type III burst could be made up of an exciter function followed by an exponential decay, as there is an observed asymmetry in the type III light-curve.  It was believed that the exciter envelope was possibly driven by the motion of the electron beam whilst the damping exponential was driven by a separate process.  Consequently, the bulk of the time profile analysis was heavily focussed in this idea, to the detriment of analysing the rise time of the radio burst profile before the peak time.

One initial theory for the damping time, and motivation for an exponential function, was that collisional damping was responsible and so plasma temperatures could be inferred \citep[see][for a discussion]{Riddle:1974aa}.  Temperatures that were similar to coronal temperatures were inferred at high frequencies.  However, studies \citep[e.g.][]{Evans:1973aa,Alvarez:1973aa,Takakura:1975aa,Poquerusse:1984aa,Ratcliffe:2014aa} found that collisional times cannot correctly explain radio burst decay times correctly, particularly near 1~AU.   

Large scale density inhomogeneity reduces the energy of the beam and must play a key role in establishing the decay profile.  As an electron beam exchanges energy with Langmuir waves during propagation, these waves are refracted to lower phase velocities by the large scale density inhomogeneity in the radially decreasing solar corona and solar wind \citep[e.g.][]{Kontar:2001ab}.  Langmuir waves are absorbed by the background plasma, depleting energy from the beam-plasma system \citep[e.g.][]{Kontar:2009aa,Reid:2013aa}.  Density inhomogeneities can suppress the generation of radio emission in the presence of significant Langmuir wave populations \citep{Ratcliffe:2014ab}.  The characteristic time for density inhomogeneity tracked the half-width half-maximum decay time of simulated type III bursts \citep{Ratcliffe:2014aa}.  The simulated time profiles were nearly symmetric around the peak time and so density inhomogeneity might not be the dominant process when the time profile is highly asymmetric.

The scattering of light from source to observer \citep{Steinberg:1971aa,Riddle:1972aa} has also been suggested to account for the exponential decay time seen in radio bursts.  Particularly at lower frequencies where the time profile is increasingly asymmetric.  Moreover, the early observations could not discriminate intrinsic and radio-wave propagation effects.  Recent imaging and spectroscopic observations of the fine structures \citep{Kontar:2017ab} demonstrated that radio-wave propagation effects, and not the properties of the intrinsic emission source, dominate the observed spatial characteristics and hence play an important role for the time profile.

The characteristic exponential decay time that was found from fits to radio bursts increases as a function of decreasing frequency \citep[e.g.][]{Aubier:1972aa,Evans:1973aa,Alvarez:1973aa,Barrow:1975aa,Poquerusse:1977aa}.  This increase can be fit with a power-law over a wide range of frequencies \citep{Alvarez:1973aa} finding a spectral index close to one.  At different frequencies, the decay time has been found to be proportional to the duration of type III bursts, using the full-width, half-maximum of the flux profile.  Later, \citet{Poquerusse:1977aa} found the same relation at a single frequency, around $169$~MHz, insinuating that both quantities are linked to the electron beam dynamics at this frequency rather than the decay constant being linked to the ambient plasma, as previous expected.  Whether this result holds true at lower frequencies, when the time profile is more asymmetrical, has not be researched.

The exciter profile of the type III bursts (rise above background till part way through the decay phase) follows the same trend, increasing as a function of decreasing frequency \citep[e.g.][]{Aubier:1972aa,Evans:1973aa,Barrow:1975aa}.  The study by \citep{Evans:1973aa} fit a linear relation as a function of decreasing frequency between $2.8-0.067$~MHz.  What the exciter times do not analyse is the rise time, the rate of increase before the peak flux, of the type III bursts.  This has not been adequately researched in the literature.  \citet{Poquerusse:1977aa} approximated the rise time at 169~MHz using the slope of the linear rise, finding it to scale with the slope of the decay, albeit with asymmetry.

The drift rate $\pdv{f}{t}$ (measured in MHz s$^{-1}$) is typically found using the peak time in each frequency channel.  A study by \citet{Alvarez:1973ab} collated numerous observations to plot how the drift rate changes as a function of frequency, finding a power-law over four orders of magnitude between $1000-0.1$~MHz, such that $\pdv{f}{t}=-0.01f^{1.84}$, where $f$ is measured in MHz and $t$ is measured in seconds.  Since then, studies \citep[e.g.][]{Achong:1975aa,Melnik:2011aa} have found drift rates that are slightly lower than predicted by \citet{Alvarez:1973ab} at frequencies above 10~MHz.

The velocity of electron beams is typically deduced using the drift rate and assumptions regarding the background density model and emission mechanism.  Typical velocities are around 0.3~c but can have a broad range from 0.1--0.5c \citep{Suzuki:1985aa,Reid:2014ab,Carley:2016aa}.  Whilst not being mono-energetic, single velocities are usually derived from radio bursts.  Electrons over a range of velocities are believed to collectively move in a beam-plasma structure \citep[e.g.][]{Ryutov:1970aa,Kontar:1998aa,MelNik:1999ab} at the average velocity of the participating electrons.  The velocity of electron beams, as derived from type III bursts, has been reported to be slowing down as the beams propagate through the heliosphere \citep{Fainberg:1972aa,Krupar:2015aa,Reiner:2015aa}.  The drift rate has been investigated using the onset time rather than the peak time observationally \citep{Fainberg:1972aa,Dulk:1987aa,Reiner:2015aa} and numerically \citep{Li:2013aa} and a faster velocity was obtained, postulating that the front of the electron beam travels with a faster velocity.  Derived electron beam velocities are also dependent upon the conditions of the background plasma, with \citet{Li:2014aa} showing how a background Kappa distribution resulted in faster derived velocities than when the background had a Maxwellian distribution.

The bandwidth of the type III bursts (the instantaneous width in frequency space) has not frequently been analysed.  A broad instrumental coverage in frequency space is required to ascertain the spectral width.  The coherent plasma emission mechanism causes the intensity of radio emission to increase as a function of decreasing frequency down to 1~MHz \citep[e.g.][]{Dulk:1998aa,Krupar:2014aa}.  One must therefore be careful how the spectral extent of the radio burst is defined.  The bandwidth of type III bursts typically decreases as a function of decreasing frequency.  An early study by \citet{Hughes:1963aa} found bandwidth of 100~MHz with a minimum frequency of 100~MHz, although they do not state exactly exactly how the the minimum and maximum frequencies are defined from the light curve.  \citet{Melnik:2011aa} reports bandwidth from 20~MHz with frequency 27~MHz down to 10~MHz at frequency 12~MHz, although again, it is not stated how the bandwidth is defined. \citet{Melnik:2011aa} also provides the relation $\Delta f=0.6f_{\rm pe}$, where $f_{\rm pe}=\sqrt{4\pi e^2m_e/n_e}$ for the bandwidth assuming fundamental emission at $f_{\rm pe}$ for a background electron density $n_e$.

In this work, we analyse type III profile finding rise, peak and decay times between $70-30$ MHz.  The LOFAR observations allow estimates for these quantities with a high degree of accuracy.  After explaining how we use the observations in Section \ref{sec:observations}, we present the temporal observations in Section \ref{sec:rdd}.  In Section \ref{sec:dfdt_bw} we show how these quantities drift through time and frequency and use the results to discus electron beam parameters in Section \ref{sec:dynamics}.  We explain how the electrons move through space in Section \ref{sec:theory}, concluding our results in Section \ref{sec:conclusion}.

\section{Observations} \label{sec:observations}

\subsection{Event List}

We used a selection of type III bursts observed by LOFAR between April--Sept 2015.  LOFAR was only observing for 2 hour intervals on specific days.  As our goal was to characterise the behaviour of individual events, we only selected isolated type III bursts.  We thus ignored type III bursts that overlapped with other radio activity, showed too much fine structure or were too faint to be detected between frequencies 70--30 MHz.  We selected 31 type III bursts that satisfied this criteria.

The LOFAR observations were taken using around $0.01$~second time resolution.  To improve signal-noise ratio, the observations were integrated to around 0.1 seconds time resolution.  The frequency range around 80--30~MHz was covered by 258 sub-bands, each with 16 channels.  Again, to improve signal-noise we integrated over each channel to use the 0.195~MHz sub-band frequency resolution.  The sub-band frequency resolution is comparable to the frequency resolution of similar instruments (e.g. Nan\c{c}ay Decametre Array) but the enhanced signal-noise resolution provides lightcurves with a small standard deviation for accurate fitting analysis.  The dynamic spectra were taken as an average over 169 dynamic spectra using the LOFAR tied-array mode \citep{Stappers:2011aa}, pointed at different positions across the solar disc \citep[e.g.][]{Morosan:2014aa,Reid:2017ac}.  Whilst we used measurements between 80--30~MHz to obtain our results, the number of radio bursts that had a high enough signal-noise ratio above 70~MHz was small.  We therefore only reported the trends in characteristic times between 70--30~MHz.

\subsection{Type III flux profile}

Type III flux profiles as a function of time and frequency provide information on the dynamics of electron beams that propagate through the solar corona and the heliosphere.  In this study, we are particularly interested in the rise, peak and decay times at individual frequencies, together with how these parameters vary as a function of frequency.

Whilst obtaining these parameters directly from the data measurements is desirable, accuracy in time is restricted to the cadence of the measurements, and fluctuations in the data from noise can cause spurious results.  Moreover, the background flux must be accounted for.  To resolve the above problems, we fit the flux as a function of time and obtained characteristic parameters from the fit.

Previous studies \citep[e.g.][]{Aubier:1972aa,Evans:1973aa,Achong:1975aa} have not characterised a rise or decay profiles but assumed an exponential decay function \citep[see e.g.][for an example]{Aubier:1972aa,Evans:1973aa}.  We make no such assumptions and use the half-width, half-maximum (HWHM) to characterise rise and decay. Whilst the exponential function is a reasonable approximation for the decay constant, it is not a good envelope for the entire flux profile.  
The peak of the type III burst time profile with a high temporal resolution typically has a rounded top that is inconsistent with exponential decay.

To assess whether an exponential or Gaussian function best characterises the type III flux profile, we first subtracted a background $B(f)$ from the type III flux profiles.  We found the background for each frequency sub-band by taking the mean flux over a period of 3-5 minutes when there was reduced activity during the observations.  The background $B(f)$ is assumed to be constant in time.  The LOFAR observations of type III burst time profile \citep{Kontar:2017ab} suggest asymmetric time profile at 30-40 MHz. Therefore, we fit the intensity of the type III bursts for each frequency using an exponential function of the asymmetric form
\begin{equation}\label{eqn:exp}
I(t)=A\exp\left(- \frac{|t-t_0|}{\tau}\right),\quad \tau=
	\begin{cases}
		\tau_{\rm r},& \text{if } t \leq t_0 \\
    	\tau_{\rm d},& \text{if } t > t_0
	\end{cases}
\end{equation}
A different characteristic rise and decay is required to capture the asymmetry in the type III flux around the peak time.  We also fit the type III bursts using a Gaussian function of the form
\begin{equation}\label{eqn:gau}
I(t)=A\exp\left(- \frac{(t-t_0)^2}{2\tau^2}\right), \quad \tau=
	\begin{cases}
		\tau_{\rm r},& \text{if } t \leq t_0 \\
    	\tau_{\rm d},& \text{if } t > t_0
	\end{cases}
\end{equation}
All fitting in this work was done using \textbf{mpfitfun} and \textbf{mpfitexy} \citep{Markwardt:2009aa} which provides an associated statistical error on each fitting parameter.

\begin{figure}\center
\includegraphics[width=\wfig,trim=100 0 0 0,clip]{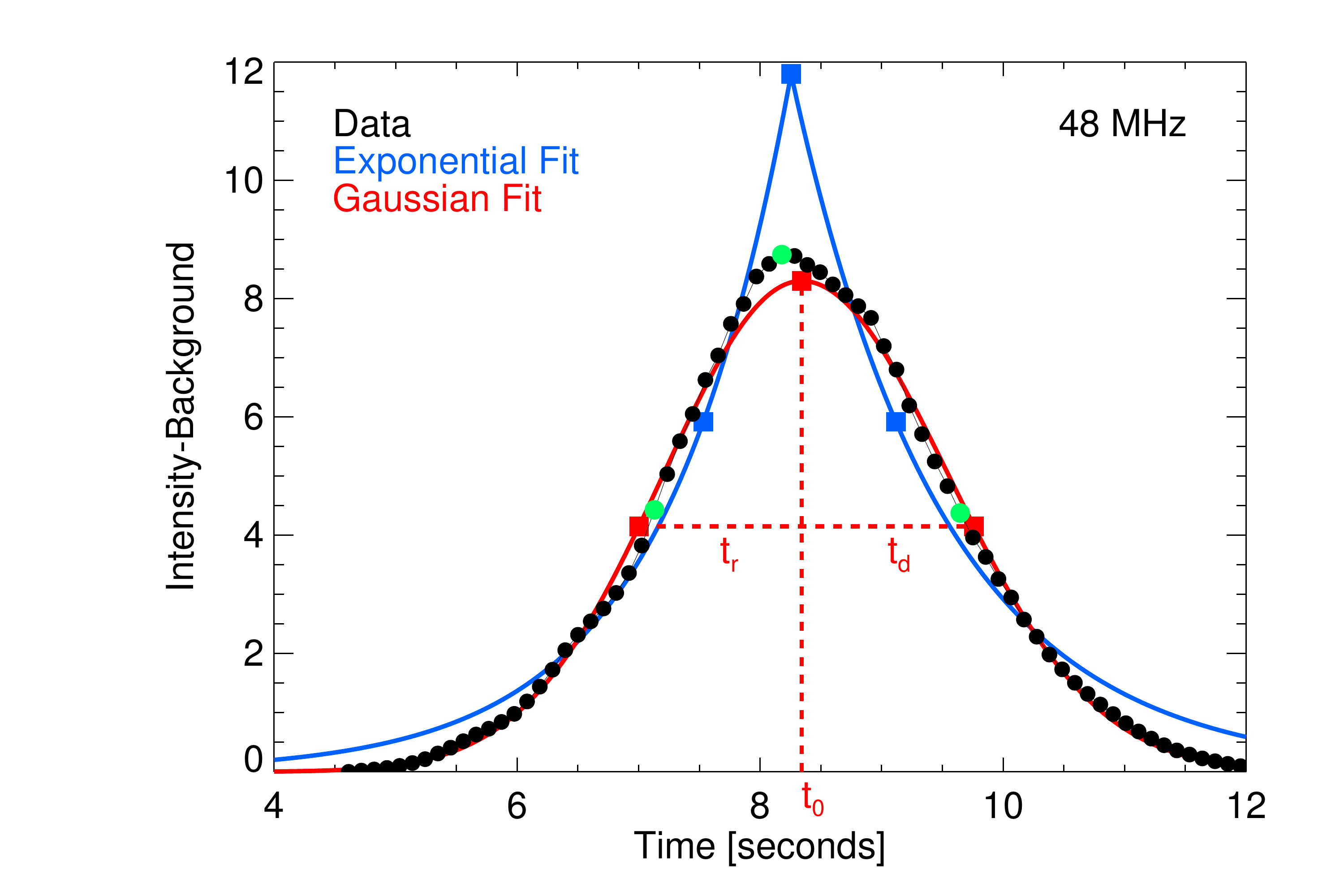}
\includegraphics[width=\wfig]{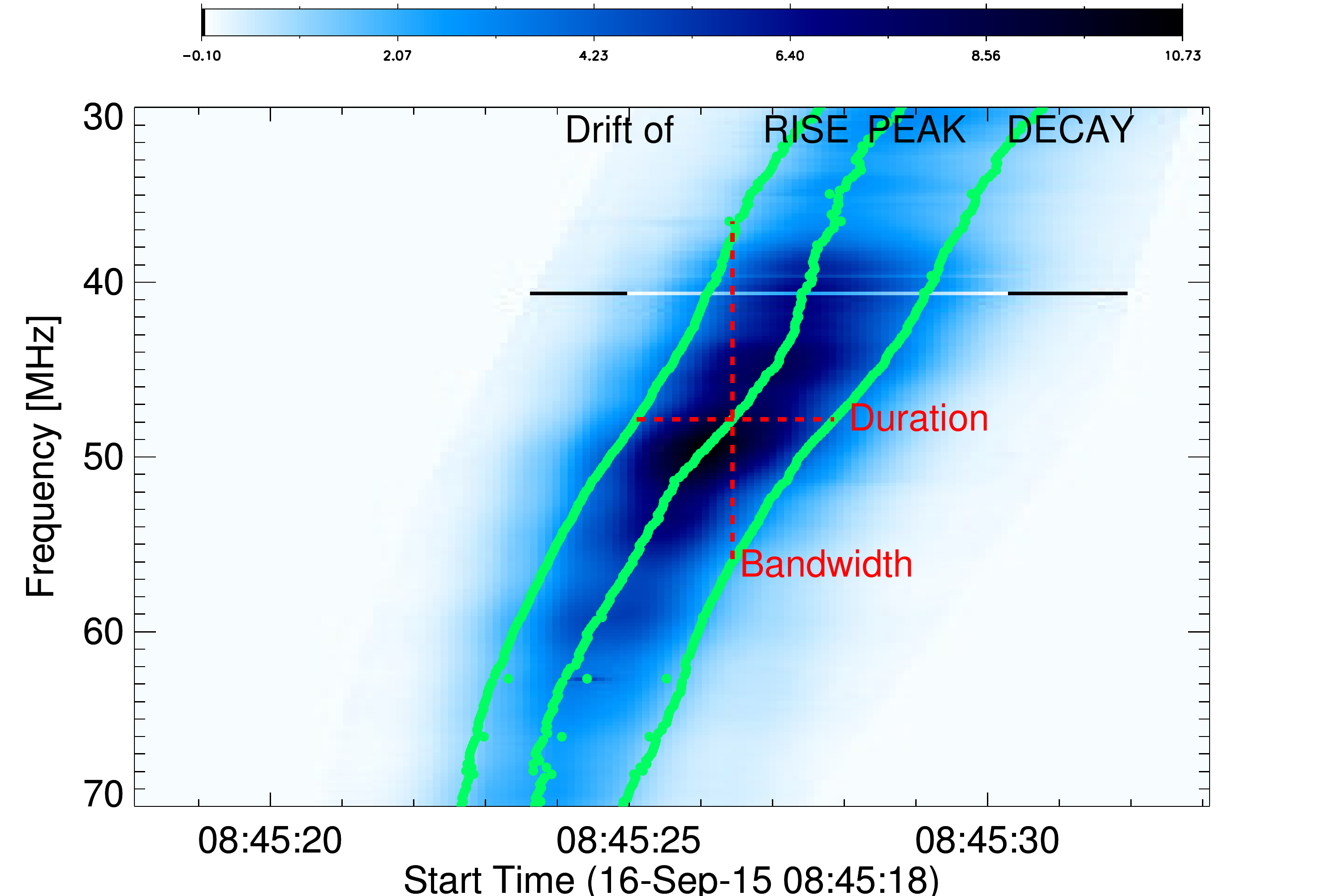}
\caption{Top: background subtracted type III flux profile around 48 MHz.  The peak flux and HWHM points are shown as green circles.  A Gaussian (red) and exponential (blue) fit are shown, together with their peak flux and HWHM.  The Gaussian fit better tracks the HWHM points in the data, and the rise, peak and decay times derived from the Gaussian fit are shown using the red dashed lines.  Bottom: dynamic spectrum of the same burst.  The drift of the rise, peak and decay as a function of frequency are shown, together with the duration and bandwidth of the radio burst around 48~MHz.} 
\label{figure:t3_flux}
\end{figure}

An example flux profile is shown in Figure \ref{figure:t3_flux} along with a Gaussian and exponential fit to the data.  The rise and decay times, $t_{\rm r}$ and $t_{\rm d}$, were characterised using the half-width, half-maximum from each fit (e.g. for the rise time, exponential $t_{\rm r} = \tau_{\rm r}\log(2)$, Gaussian $t_{\rm r} = \tau_{\rm r}\sqrt{2\log(2)}$).  We also found the HWHM directly from the data using the point farthest from the peak with a flux at least 50\% of the peak flux.  A comparison found that the Gaussian fit was substantially closer to the HWHM found from the data.  The exponential function typically underestimated the rise and decay times, shown by Figure \ref{figure:t3_flux}.  There were a few radio bursts where the decay time from the exponential fit better agreed with the HWHM found from the data.  These bursts typically had a larger asymmetry, with the decay times being notably larger than the rise times.  However, for consistency we used a Gaussian fit for all the radio burst time profiles.

Characterising the rise, peak and decay times allows us to see how each characteristic time drifts as a function of frequency.  Figure \ref{figure:t3_flux} shows how these times result in three different drift rates for a single type III burst.  The duration $t_{\rm D}$, found from the FWHM, is also indicated around 48~MHz.

The instantaneous bandwidth of the type III radio burst, the width of the type III burst in frequency space, can be found at each point in time.  There is no standard definition as to how the instantaneous bandwidth is found in the literature.  We introduce here a technique to define the bandwidth using the type III rise and decay times.  At each point in time we find the highest type III frequency by finding the highest frequency that has not stopped emitting using $t_0+t_{\rm d}$.  Similarly we find the lowest type III frequency by finding the lowest frequency that has started emitting using $t_0-t_{\rm r}$.  The bandwidth is defined as the difference between the highest and lowest frequency.

For the results in this study, we therefore use the Gaussian fit described by Equation \ref{eqn:gau} to obtain the dynamic spectrum parameters of interest. The following type III parameters are found from the fits:
\begin{itemize}
\item The rise time $t_{\rm r}$ is found at $t<t_0$ from HWHM, $\tau_r\sqrt{2\log(2)}$.
\item The decay time $t_{\rm d}$ is found at $t>t_0$ from the HWHM $\tau_d\sqrt{2\log(2)}$.
\item The duration $t_{\rm D}$ is found using the full-width half-maximum (FWHM).
\item The drift rate is found from the change in rise, peak and decay time as a function of frequency.
\item The bandwidth is found from the frequency width between the HWHM at different frequencies.
\end{itemize}

\section{Type III rise, decay and duration} \label{sec:rdd}

\subsection{Rise time}

\begin{figure}\center
\includegraphics[width=\wfig,trim=100 0 0 0,clip]{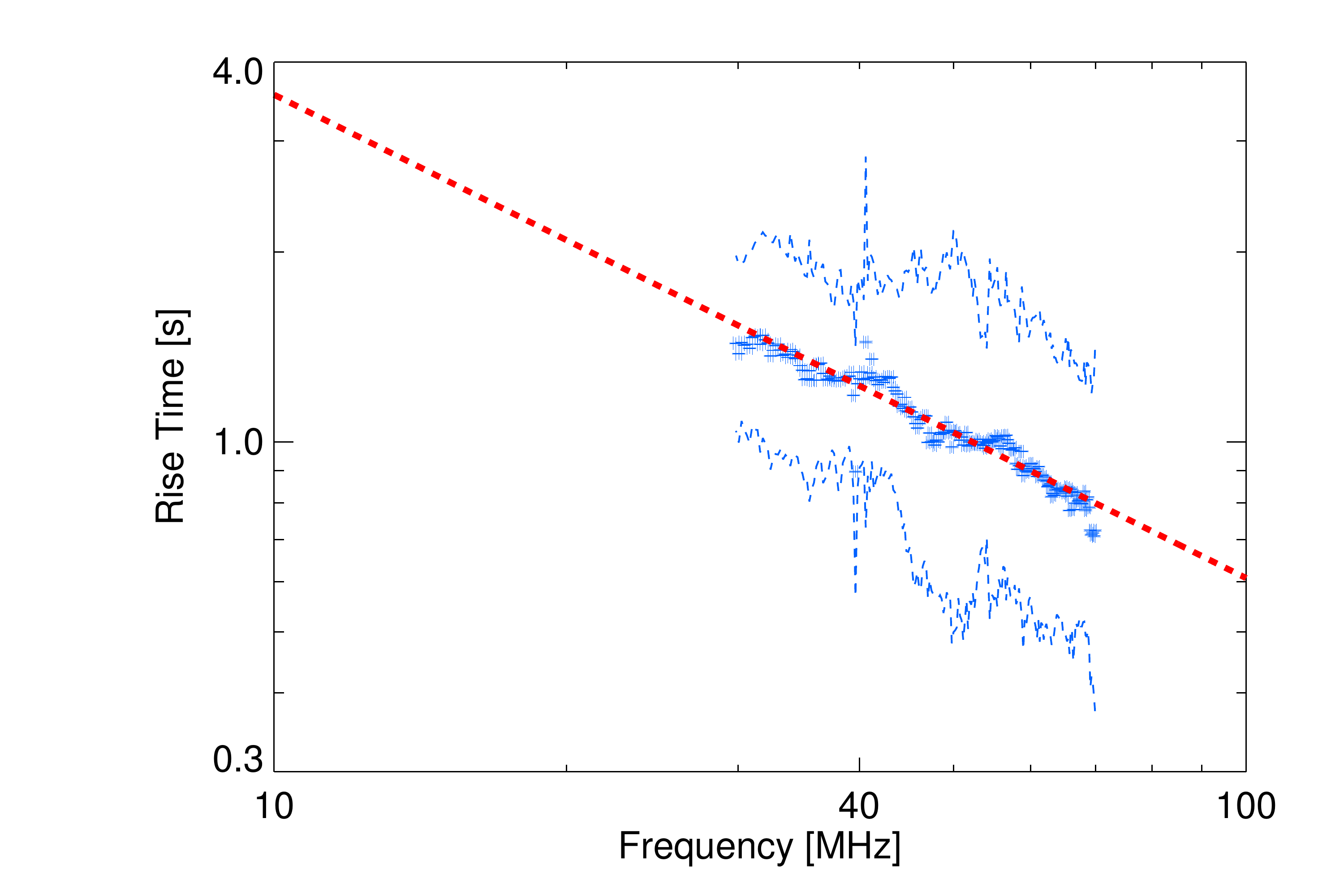}
\includegraphics[width=\wfig,trim=100 0 0 0,clip]{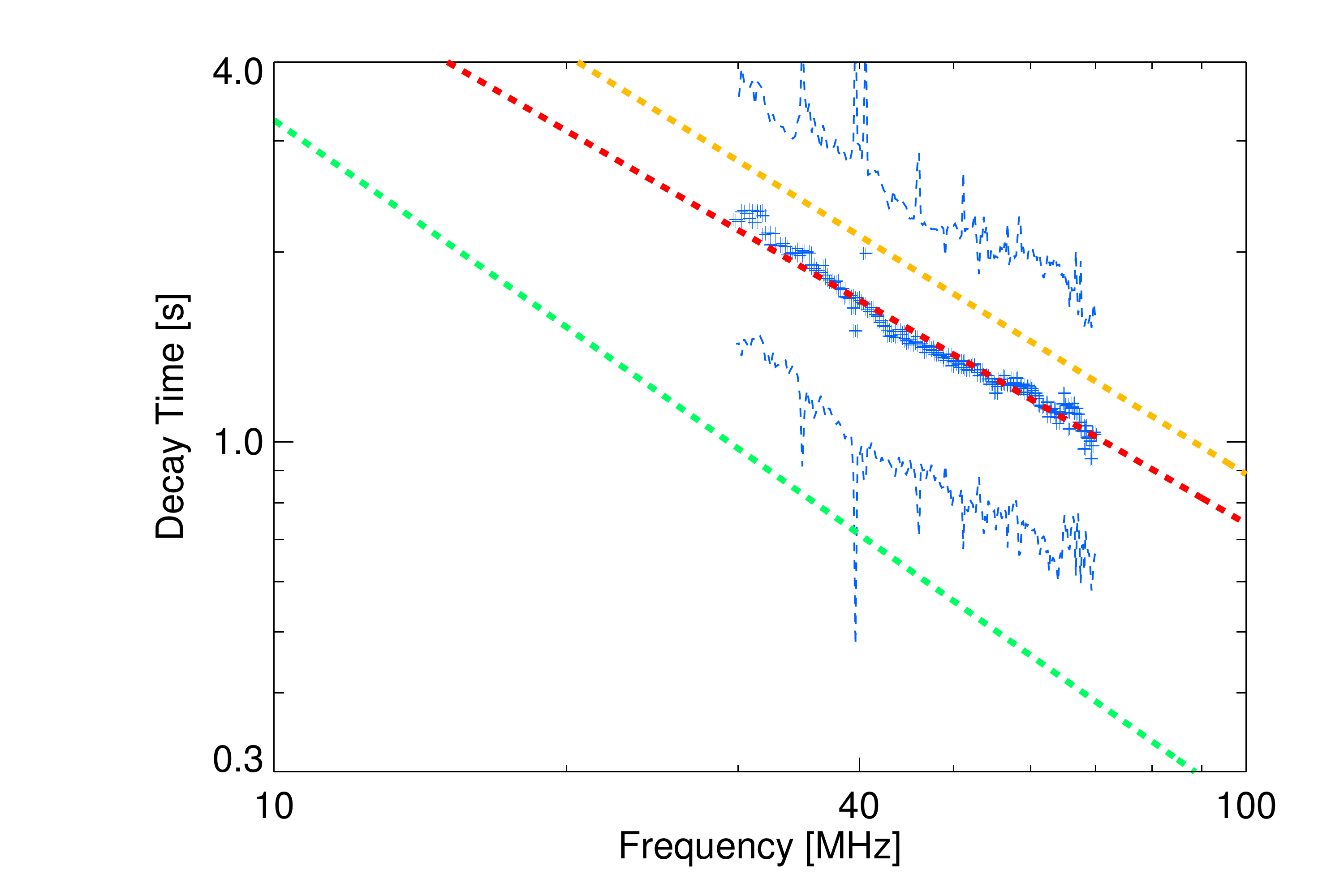}
\caption{Mean rise time (top) and decay time (bottom) as a function of frequency for all the analysed type III bursts, as found from the HWHM, $t_{\rm r}$ and $t_{\rm d}$, respectively.  The top and bottom blue dashed lines show the spread of the data, found from the standard deviation.  The red dashed line show the fits $t_{\rm r} = (1.5\pm0.1)~f^{-0.77\pm0.14}$ and $t_{\rm d} = (2.2\pm0.2)~f^{-0.89\pm0.15}$ for frequency (per 30~MHz).  For the decay time, the green and orange dashed lines show the fits from \citet{Evans:1973aa} and \citet{Alvarez:1973aa}, respectively, assuming a HWHM of $t_{\rm d}=\tau_d\log(2)$, from the derived exponential profiles $\tau_d$.}
\label{figure:rise_decay_time}
\end{figure}

The rise time, $t_{\rm r}$, of the type III bursts for each frequency is determined from the HWHM of the flux before $t_0$, the time of peak flux.  Figure \ref{figure:rise_decay_time} shows how the mean rise time varies as a function of frequency over all the type III bursts.  The mean is calculated from a weighted function of each radio burst such that
\begin{equation}\label{eqn:mean_time}
\bar{t_{\rm r}} = \frac{\sum_{i=1}^N (t_{\rm{r},i}\sigma_i^{-2})}{\sum_{i=1}^N (\sigma_i^{-2})}.
\end{equation}
where $\sigma_i^2$ is the variance associated with $t_{\rm{r},i}$, obtained from the fitting of the type III burst profile using \textbf{mpfit}, and the weighted function is $w_i=1/\sigma_i^2$.

The mean rise time increases as a function of decreasing frequency.  We have fitted the change in rise time using a power-law function such that 

% \begin{equation}
% t_{\rm r} = A \left(\frac{f}{[30~\rm{MHz}]}\right)^b
% \end{equation}
\begin{equation}
t_{\rm r} = (1.5\pm0.1) \left(\frac{f}{30~\rm{MHz}}\right)^{-0.77\pm0.14}.
\end{equation}
The error on the rise time for each individual burst is very small but there is a significant spread in the rise times over all radio bursts, indicated by the standard deviation in the rise times in Figure \ref{figure:rise_decay_time}.  Consequently, we used the standard deviation of the frequency and rise times as an error for the fit to account for this large range in rise times.

Whilst the mean rise time increases as a function of decreasing frequency, the rise time does not increase in such a smooth manner for individual type III bursts.  The bursty nature in the dynamic spectrum of type IIIs causes the rise time in some frequency ranges to have constant rise times or steeply increasing rise times or even decreasing rise times.

\subsection{Decay time}

The decay time, $t_{\rm d}$, for each frequency is determined from the HWHM of the flux after $t_0$, the time of peak flux.  Figure \ref{figure:rise_decay_time} shows how the mean decay time varies as a function of frequency over all type III radio bursts.  The mean is calculated from a weighted function of each radio burst using Equation \ref{eqn:mean_time} using $t_{\rm{d},i}$ instead of $t_{\rm{r},i}$.  The spread in the decay times is indicated by the standard deviation in the decay times in Figure \ref{figure:rise_decay_time}.

% \begin{figure}\center
% \caption{Mean decay time as a function of frequency for all the type III bursts, as found from the HWHM, $t_d$.  The red dashed line shows the log-log fit between 30--70 MHz.  The blue dashed lines show the spread of the data, found from the standard deviation.  The green and orange dashed lines show the fits from \citet{Evans:1973aa} and \citet{Alvarez:1973aa}, respectively, assuming a HWHM of $0.69\tau$, from the derived exponential profiles $\tau$.}
% \label{figure:decay_time}
% \end{figure}

We have fitted the decay time using a power-law function such that 
\begin{equation}
t_{\rm d} = (2.2\pm0.2) \left(\frac{f}{30~\rm{MHz}}\right)^{-0.89\pm0.15}.
\end{equation}
We used the standard deviation of the frequency and decay time as an error for the fit.  The spectral index $b$ captures the general trend that the decay time increases as a function of decreasing frequency.  The decay times are slightly lower than the decay times, found from numerical simulations of a type III burst \citep{Ratcliffe:2014aa}, which had a decay of 1 second around 100~MHz and 8 seconds around 10~MHz.

We can compare our results to the previous fit found from \citet{Alvarez:1973aa} who used an exponential function to obtain a characteristic decay constant $\tau_d$.  They fit data from multiple studies between the frequencies 0.06--200~MHz (only two data points below 0.2~MHz), finding $\tau_{d} = 4f^{-0.95}$ for frequency $f$ (per 30~MHz).  We scaled $\tau_d$ to the HWHM using $t_{\rm d} = \tau_d\log({2})$.  The corresponding fit is over-plotted in Figure \ref{figure:rise_decay_time}.  The slope agrees well with our observations, being within one standard deviation, but the absolute magnitude is higher.

The characteristic decay constant $\tau_d$ for type III bursts was also analysed by \citet{Evans:1973aa} but between the frequencies $2.8$ -- $0.067$~MHz.  They used a similar exponential fit, finding $\tau_{d} = (1.4^{+4.9}_{-1.2})f^{-1.09\pm0.05}$ for frequency $f$ (per 30~MHz).  Again, we scaled $\tau_d$ to the HWHM using $t_{\rm d} = \tau_d\log({2})$.  Over-plotting the fit on Figure \ref{figure:rise_decay_time} shows that extrapolating the fit from lower frequencies back to higher frequencies does not obtain an accurate prediction.  The comparison suggests that a single power-law might not capture the increase in the decay rate as a function of decreasing frequency from the corona (e.g. 50~MHz) all the way out into interplanetary space (e.g. 50~kHz).

\subsection{Asymmetry in the time profile}

\begin{figure}\center
\includegraphics[width=\wfig,trim=60 0 0 0,clip]{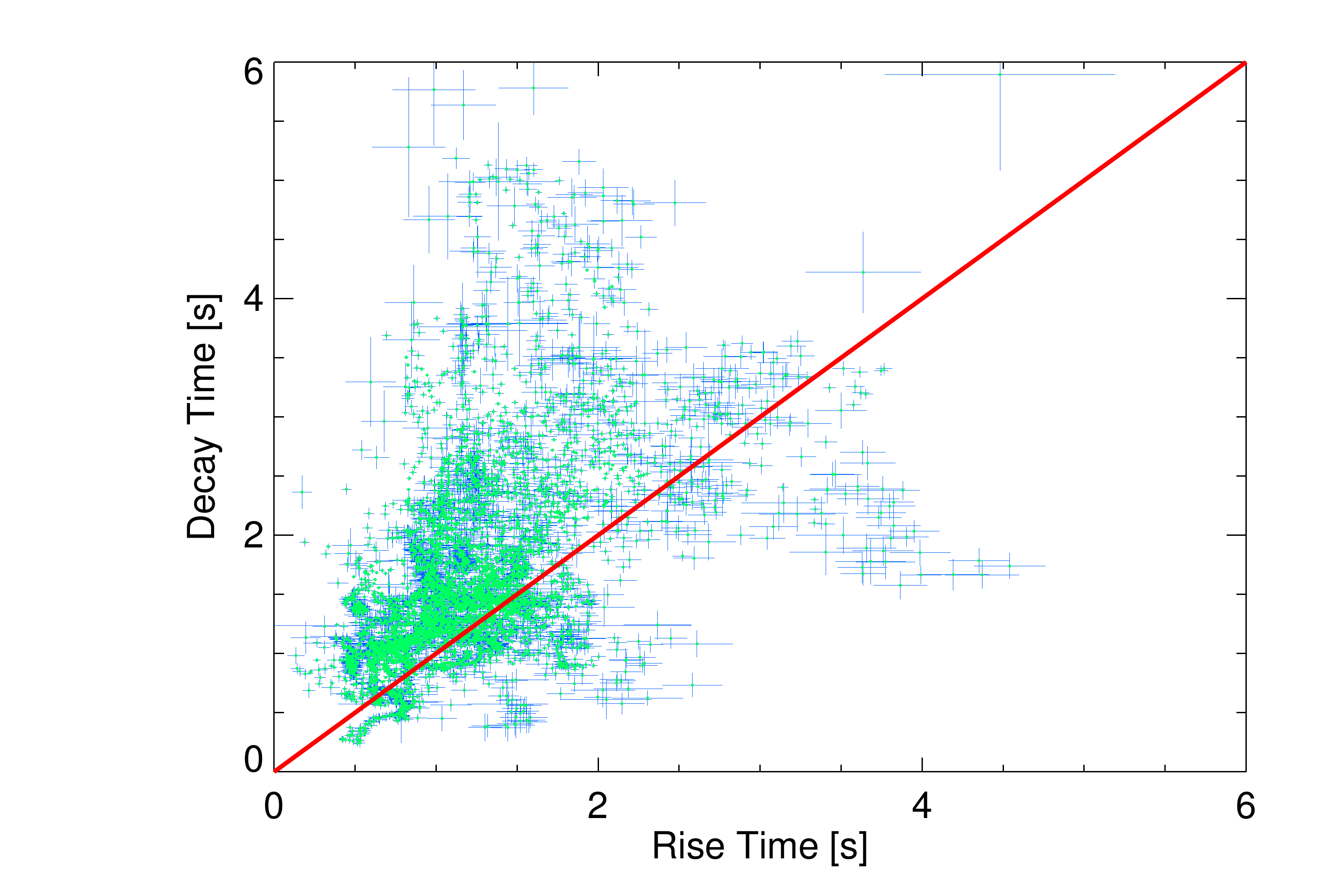}
\includegraphics[width=\wfig,trim=60 0 0 0,clip]{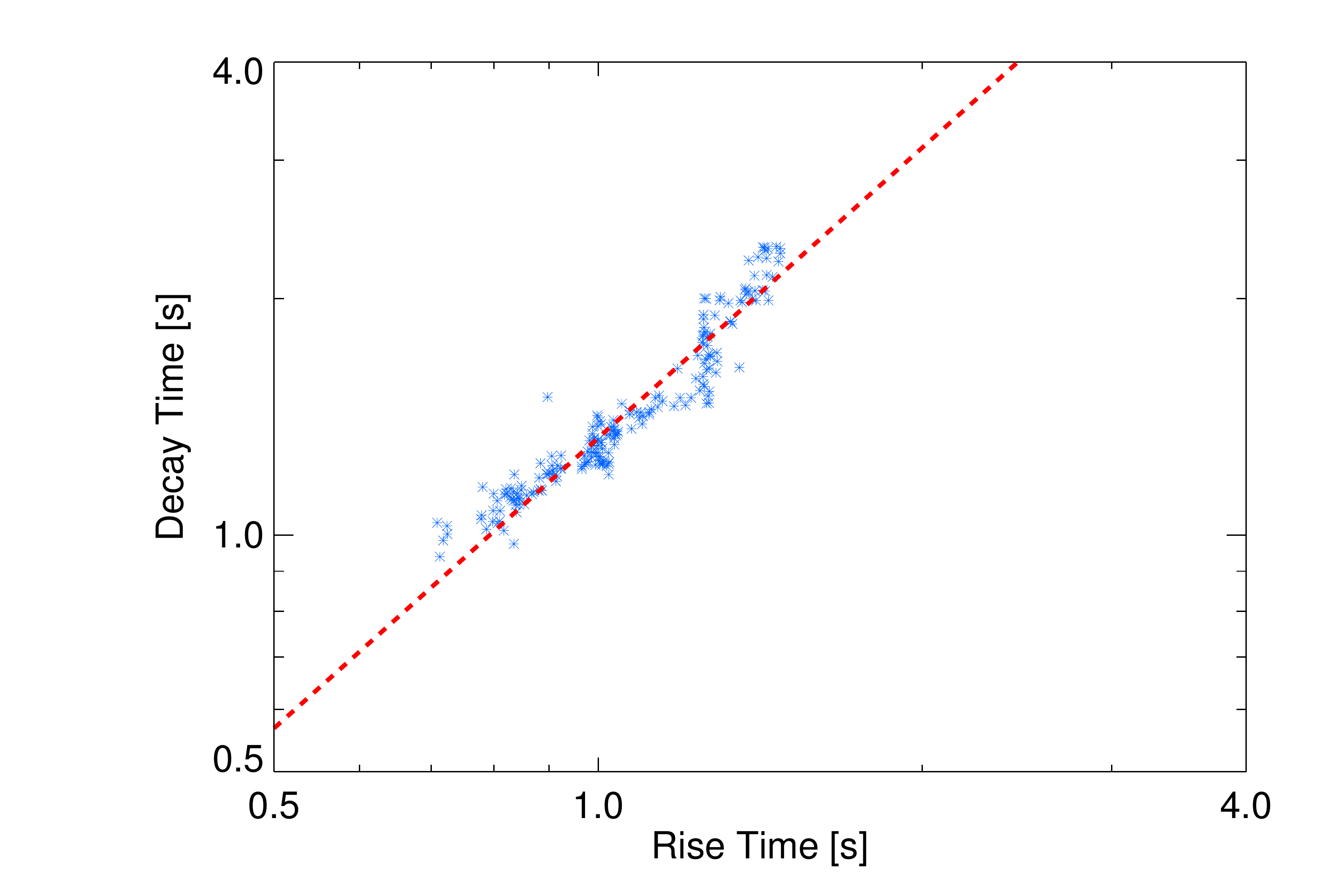}
\includegraphics[width=\wfig,trim=60 0 0 0,clip]{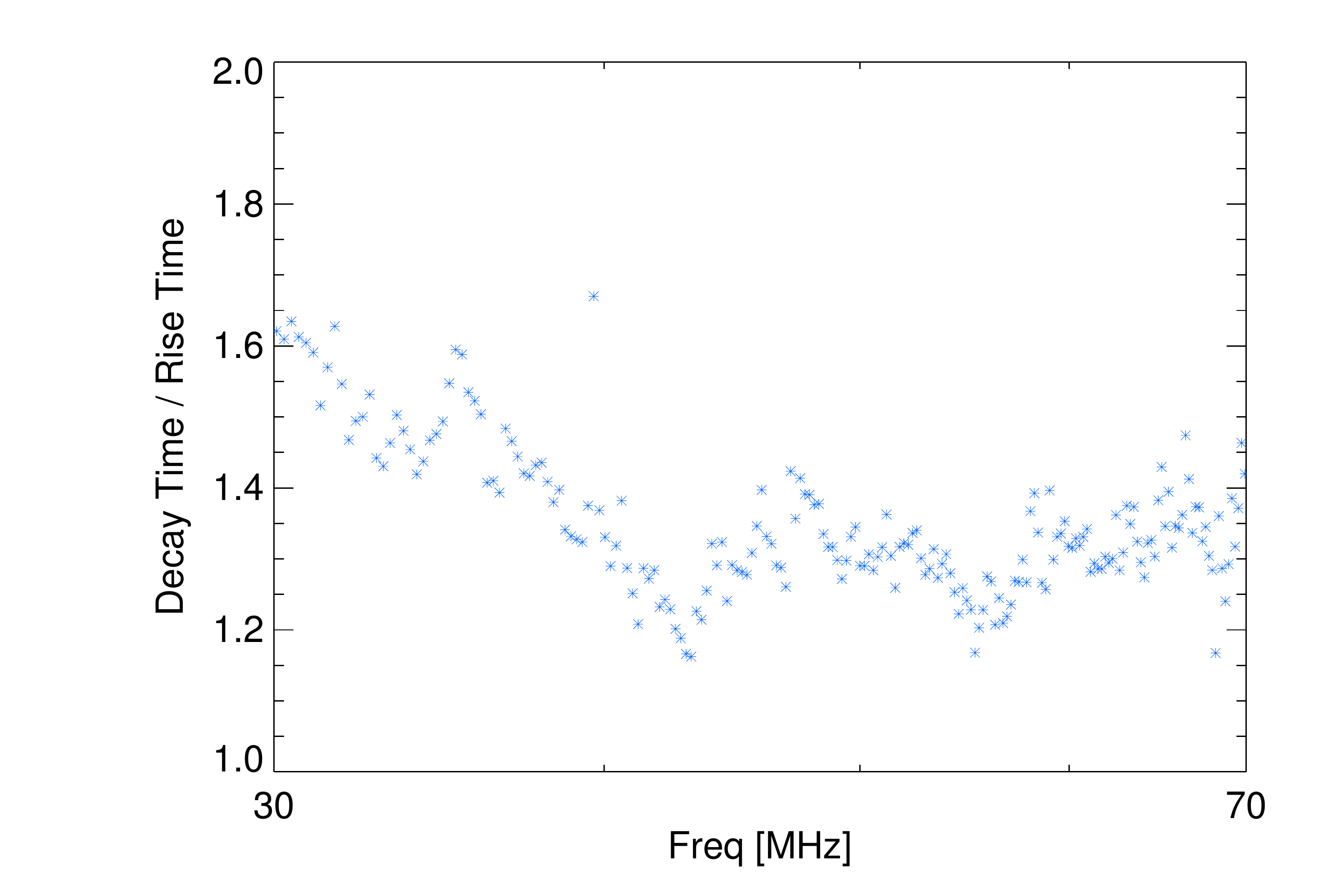}

\caption{Top: asymmetry of the type III burst time profile along with the errors (in blue) on the rise and decay times.  The red line highlights $y=x$.  The higher decay times than rise times is clear, with 77\% of the time profiles having a longer decay time than rise time.  Middle: the mean rise time vs mean decay time including the power-law fit to the data $t_{\rm r} = (1.33\pm0.07)~t_{\rm d}^{1.23\pm0.27}$ for time [seconds].  Bottom: the ratio of decay time over rise time as a function of frequency.}
\label{figure:asy_time}
\end{figure}

Type III time profiles typically show asymmetry, with the decay time being larger than the rise time.  Figure \ref{figure:asy_time} shows the rise time plotted against the decay time along with the associated errors from the fitting.  There is a clear asymmetry, with the decay time being higher than the rise time in 77\% of the time profiles.  The Pearson correlation coefficient is 0.51 and does not change much when considering the asymmetry of rise and decay time at individual frequencies.

A correlation has been shown between the mean duration and the mean decay constant for type III bursts as a function of frequency \citep[e.g.][]{Aubier:1972aa,Barrow:1975aa,Poquerusse:1977aa,Abrami:1990aa}.  In a similar vein, Figure \ref{figure:asy_time} shows that the mean rise time and mean decay time are correlated, with a Pearson correlation coefficient of 0.95.  We have fitted the ratio using a power-law function such that
\begin{equation}
t_{\rm r}=(1.33\pm0.07)\left(\frac{t_{\rm d}}{1~\rm{Second}}\right)^{1.23\pm0.27}
\end{equation}
using the standard deviation of both variables as an error for the fit.  Figure \ref{figure:asy_time} also shows the general trend of an increase in the ratio of decay time over rise time as the frequency decreases.

\subsection{Duration}

\begin{figure}\center
\includegraphics[width=\wfig,trim=60 0 0 0,clip]{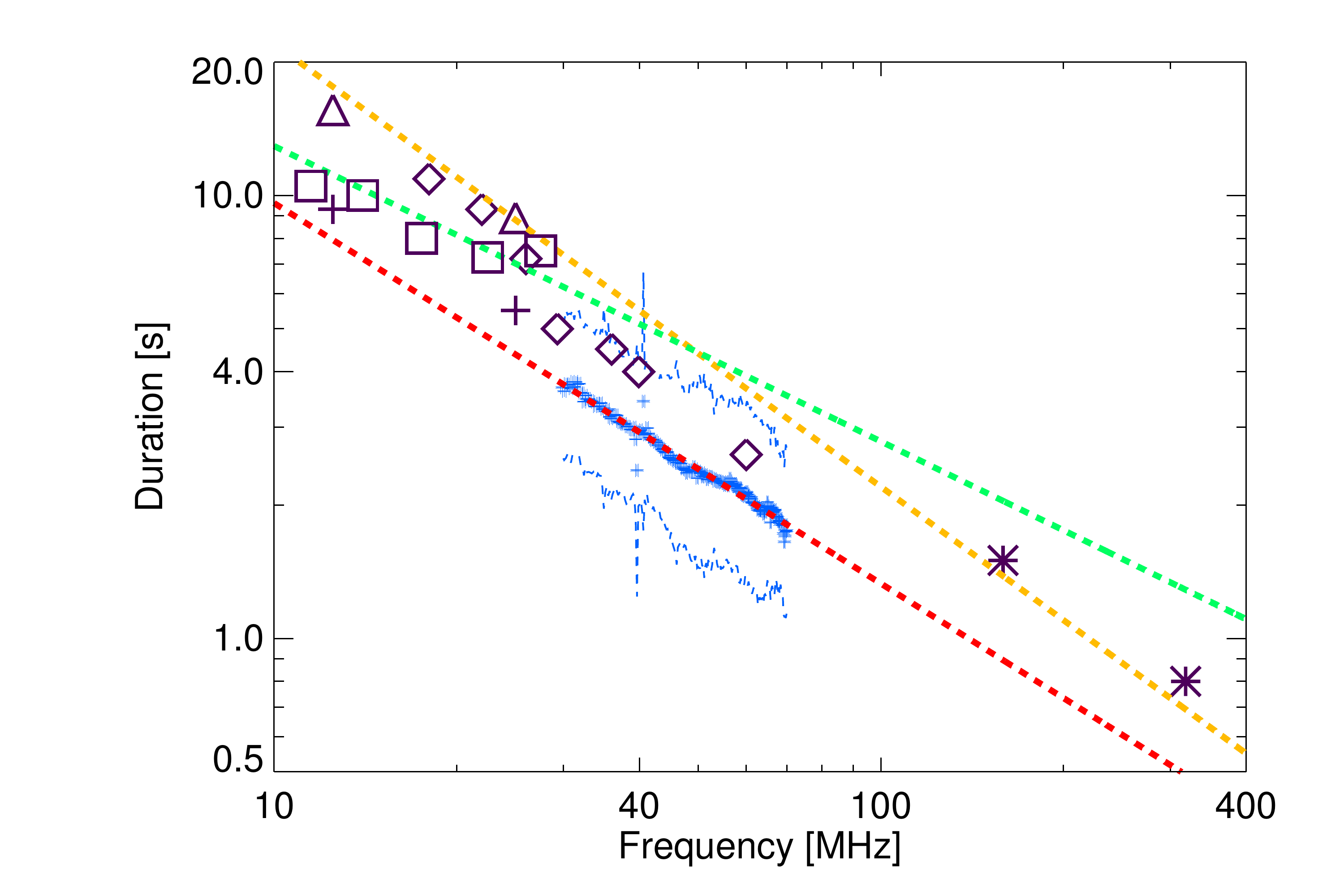}
\caption{Mean duration as a function of frequency for all the type III bursts, as found from the FWHM.  The blue dashed lines show the spread of the data from the standard deviation.  The red dashed line shows the fit $t_{\rm D} = (3.7\pm0.2)~f^{-0.86\pm0.11}$ for frequency (per 30~MHz).  The purple diamonds show the FWHM found from \citet{Aubier:1972aa} and \citet{Barrow:1975aa}.  The purple pluses and triangles show the FWHM for fundamental and harmonic bursts, respectively, from \citet{Tsybko:1989aa}.  The purple squares show the FWHM for powerful ($>1000$~SFU) type III bursts, from \citet{Melnik:2011aa}.  The orange and green dashed lines are, respectively, the fit in frequency (per 30~MHz) of $t_{\rm D}=7.3f^{-1}$ reported by \citet{Suzuki:1985aa} and $t_{\rm D}=6.2f^{-2/3}$ reported by \citet{Elgaroy:1972aa}.}
\label{figure:duration}
\end{figure}

We classify the type III duration, $t_D$, as the full-width half-maximum (FWHM) of the type III bursts.  Figure \ref{figure:duration} shows how the duration varies as a function of frequency over all the type III bursts, including the corresponding standard deviations.  We have fitted the duration using a power-law function such that 
\begin{equation}
t_{\rm D} = (3.7\pm0.2) \left(\frac{f}{30~\rm{MHz}}\right)^{-0.86\pm0.11}.
\end{equation}
The standard deviation of frequency and duration have been used to calculate the errors.  The power-law fit shows the general trend of an increasing duration as a function of decreasing frequency.

To compare with other observations, we have over-plotted the FWHM as derived by \citet{Aubier:1972aa,Barrow:1975aa,Tsybko:1989aa,Melnik:2011aa} at similar frequencies.  We also show the fits for duration as a function of frequency (per 30~MHz) of $t_{\rm D}=7.3f^{-1}$, reported by \citet{Suzuki:1985aa} and $t_{\rm D}=6.2f^{-2/3}$, reported by \citet{Elgaroy:1972aa}.  However, we note that the lower frequency durations involved in the fit by \citet{Elgaroy:1972aa} were taken from \citet{Alexander:1969aa} who calculated decay times, not durations.   As such, it underestimates the duration at the low frequencies which might explain the smaller exponent.

We find that previously reported mean durations are noticeably higher.  Nevertheless, the standard deviations show a large spread in the duration of the different type III bursts, with past observations being within one standard deviation from our observational mean.  Again, the large standard deviations in the duration highlight the spread in values over all the type III radio bursts.  The duration of type III bursts must be dependent upon the properties of the electron beams that generated the radio bursts, and perhaps the background plasma properties.  This is discussed in detail in Section \ref{sec:theory}.

% Li et al 2009 second paper has comprable durations....

\section{Type III bandwidth and frequency drift} \label{sec:dfdt_bw}

\subsection{Bandwidth}

The instantaneous bandwidth of a type III radio burst, the width of the type III burst in frequency space, can be found when the highest and lowest frequencies are observed in the dynamic spectrum.  As explained in Section \ref{sec:observations}, the bandwidth is defined by the frequency difference between the highest frequency that has not stopped emitting (using $t_0+t_{\rm d}$) and the lowest frequency that has started emitting (using $t_0-t_{\rm r}$).  For each time $t$, if there are no frequencies where either $t_0-t_{\rm d}$ or $t_0+t_{\rm r}$ can be found from the dynamic spectrum then the bandwidth is not defined.

\begin{figure}\center
\includegraphics[width=\wfig,trim=100 0 0 0,clip]{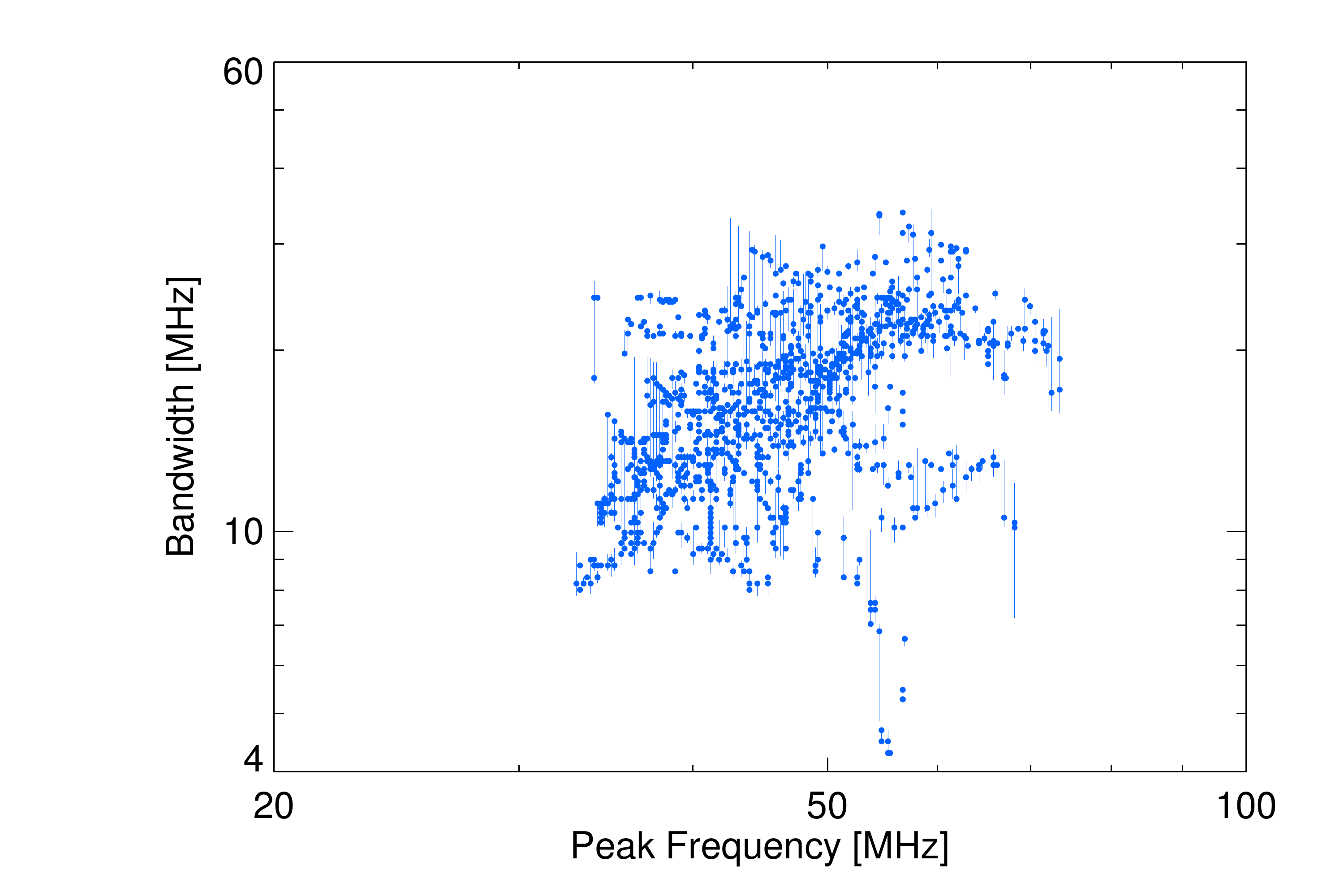}
\caption{Bandwidth $\Delta f$ of the type III bursts as a function of peak frequency.  }
\label{figure:bandwidth}
\end{figure}

Figure \ref{figure:bandwidth} shows how the bandwidth varies as a function of frequency for all the type III bursts.  The bandwidths are lower than what is reported by \citet{Hughes:1963aa,Melnik:2011aa}.  However, this is likely related to how the bandwidth is defined.  For example, \citet{Hughes:1963aa} used the minimum frequency to report the bandwidth whilst we use the central frequency.

From numerical simulations, \citet{Ratcliffe:2014aa} report the bandwidth of a type III burst, also finding slightly higher bandwidths, with a relation of $\Delta f=0.57f$ at frequencies 20--100~MHz, slightly decreasing at lower frequencies.  Again, the bandwidth is found in a slightly different way, using the FWHM in frequency space.  We refrained from using this method as it will be affected by the reported type III burst tendency for the radio flux to increase with decreasing frequencies till 1~MHz \citep[e.g.][]{Krupar:2014aa}, creating a skewed distribution.  \citet{Ratcliffe:2014aa} fit the ratio $\Delta f/f$ and found a roughly constant value of $\Delta f/f=0.57$ above 30 MHz.  We found a slightly lower relation of $\Delta f/f=0.44-0.0017f$ for frequency (per MHz).  However, the scatter on our bandwidth measurements is quite significant and the small, frequency dependent component is likely related to the reduced observations around 65~MHz.

\subsection{Drift rate}

The type III frequency drift rate $\pdv{f}{t}$ is typically measured using the peak of the radio time profile.  Characterising the time profile of the type III bursts allows us to find the drift rate not only for the peak time of each radio burst, but also for the rise time and the decay time using the HWHM.  The drift rate is known to vary as a function of frequency \citep{Alvarez:1973ab}.  The increased spectroscopic resolution of LOFAR allows the drift rate to be approximated in small bands.  We assumed a constant drift rate over a frequency width of $3$~MHz (15 sub-bands).  The drift rate over each $3$~MHz band was then calculated from a straight line fit $f = \pdv{f}{t_0} t_0+c$, together with the 1-sigma error on the fit.

\begin{figure}\center
\includegraphics[width=\wfig,trim=70 0 0 0,clip]{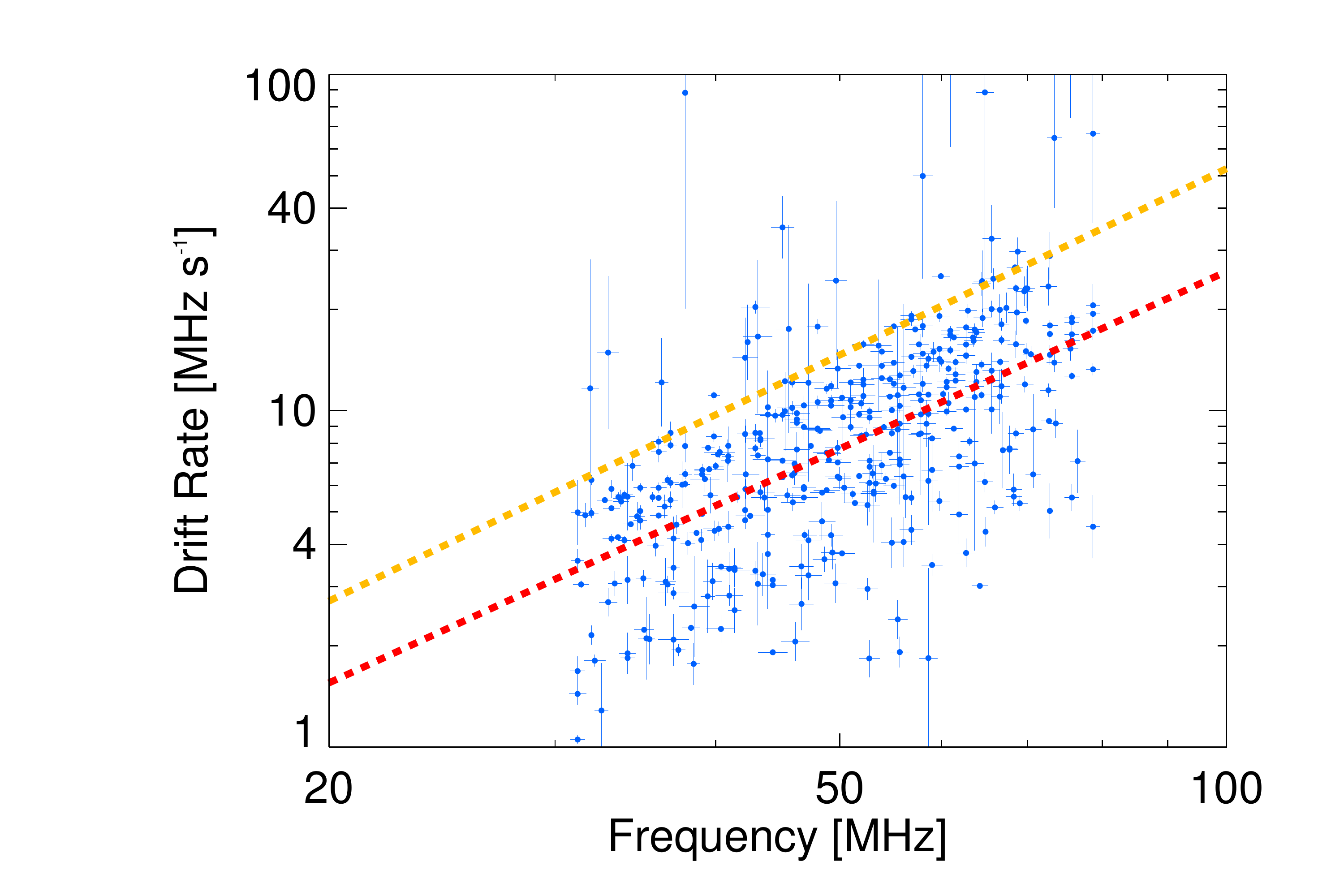}
\includegraphics[width=\wfig,trim=70 0 0 0,clip]{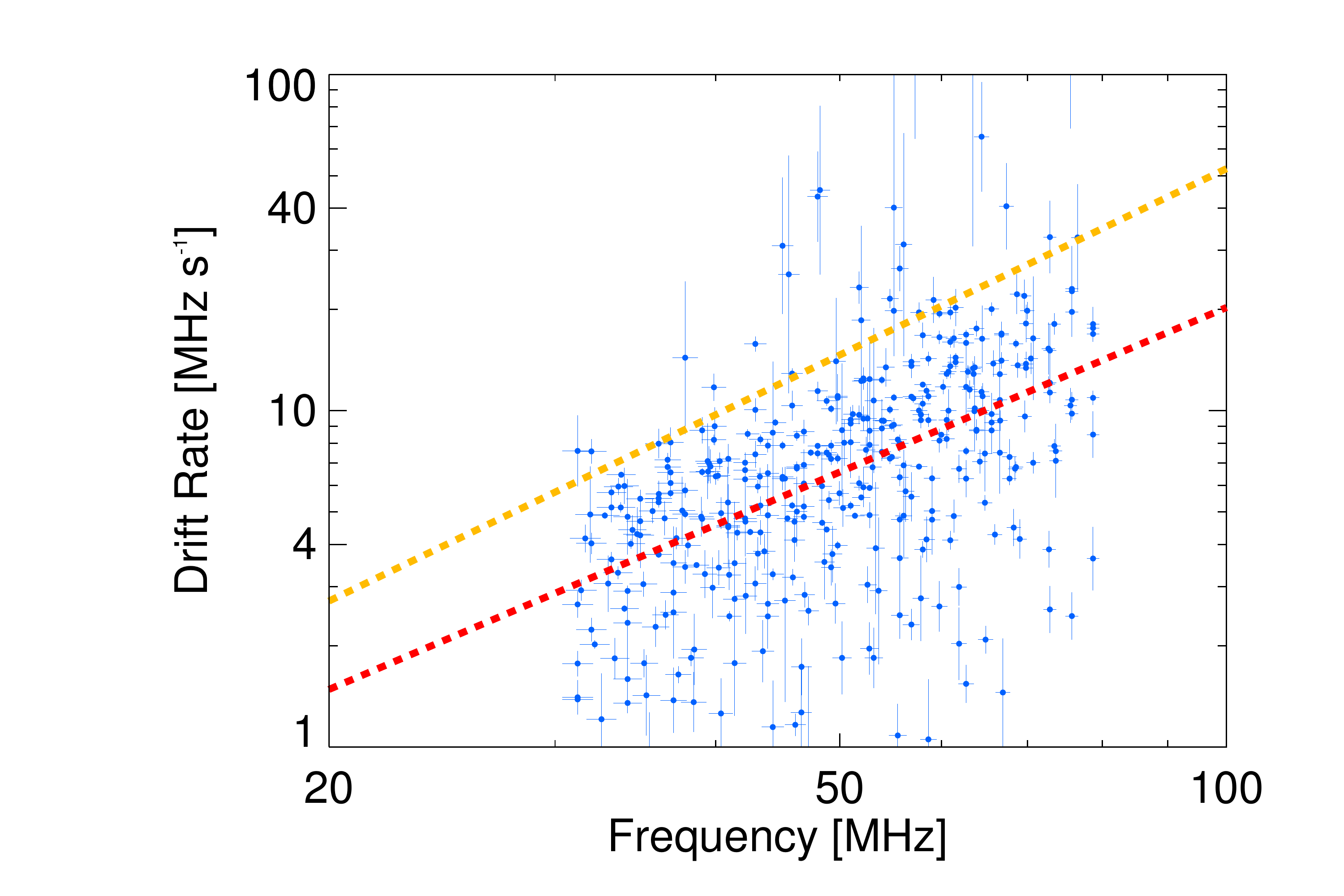}
\includegraphics[width=\wfig,trim=70 0 0 0,clip]{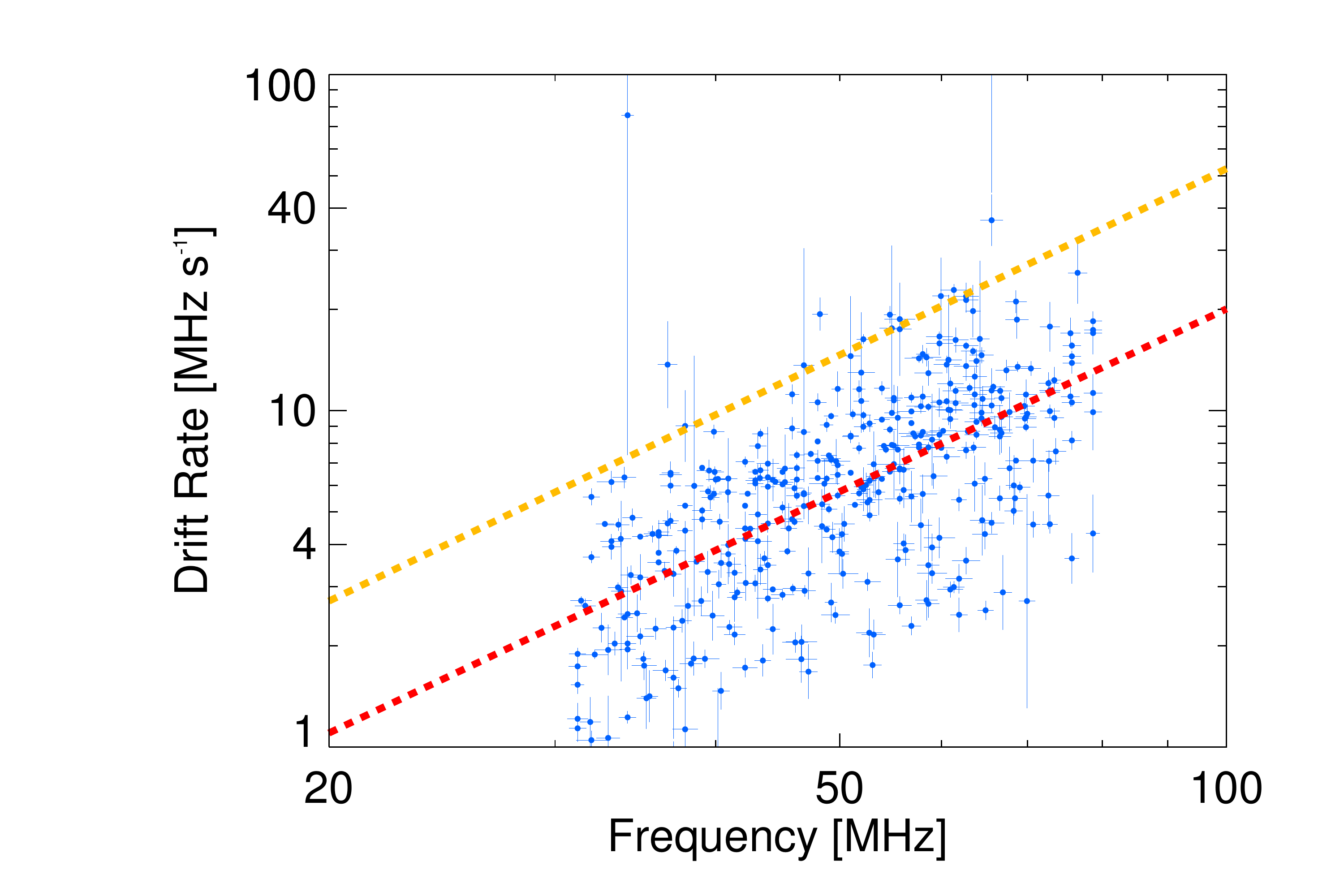}
\caption{Drift rate as a function of frequency for all type III bursts for the rise (top), peak (middle) and decay (bottom) times.  The red dashed line shows the fitted power-law to the data.  The orange dashed line shows the fit from \citep{Alvarez:1973ab}.  The reduced one-sigma errors are obtained from the fitting uncertainties in the drift rate, taking into account the fitting uncertainties in $t_0$.}
\label{figure:drift_rate}
\end{figure}

Figure \ref{figure:drift_rate} shows how the drift rate changed as a function of frequency, where the frequency is found from the central frequency from the 15 sub-bands, and the error on the fit is the corresponding standard deviation.  A power-law fit to the data is shown using
\begin{equation}\label{eqn:fit_alvarez}
% \pdv{f}{t} = -Af^{\alpha},
\pdv{f}{t} = -A \left(\frac{f}{30~\rm{MHz}}\right)^{\alpha}.
\end{equation}
The fit obtained from \citet{Alvarez:1973ab} from frequency (per 30~MHz) is $\pdv{f}{t} = -5.2f^{-1.84}$ and is shown for comparison in Figure \ref{figure:drift_rate}.  The drift rates for the rise, peak and decay times all have a noticeably lower magnitude than those observed by \citet{Alvarez:1973ab}.  These lower magnitudes agrees with the type III drift rates found by \citet{Achong:1975aa} between 18--36 MHz and the drift rates found by \citet{Melnik:2011aa} between 10--30 MHz for powerful ($>1000$~SFU) type III bursts.

\vspace{20pt}
\begin{center}
\begin{table}
\centering
\caption{Type III drift rate fit parameters for Equation \ref{eqn:fit_alvarez} using the rise, peak and decay times.}
\begin{tabular}{ c  c  c}

\hline\hline

Time 	& $A$ 							& $\alpha$ \\ \hline
% rise 	& $0.008^{+0.005}_{-0.003}$		& $1.75\pm0.13$ \\
% peak 	& $0.01^{+0.007}_{-0.004}$		& $1.67\pm0.14$ \\
% decay 	& $0.005^{+0.003}_{-0.002}$		& $1.82\pm0.12$ \\
rise 	& $3.1\pm0.2$		& $-1.75\pm0.11$ \\
peak 	& $3.8\pm0.2$		& $-1.63\pm0.13$ \\
decay 	& $1.9\pm0.1$		& $-1.80\pm0.11$ \\

\hline
\end{tabular}
\vspace{20pt}
\label{tab:drift_rate}
\end{table}
\end{center}

The drift rates associated with the rise times are marginally larger than those of the peak and larger still than the decay times.  The spectral index is slightly smaller for all three times, but is within at least two standard deviations of the slope found by \citet{Alvarez:1973ab}.

The drift rates from the peak times are very similar to the drift rates obtained from peak times in the numerical simulations by \citet{Ratcliffe:2014aa}, with the power-law fit having almost identical parameters.  The spread in the observed drift rates is too large to make any meaningful conclusions about whether multiple power-laws over different frequency ranges, used by \citet{Ratcliffe:2014aa}, would be a better fit rather than a single power-law fit.  The drift rates are also similar to simulated type III bursts at 85~MHz and $>100~MHz$ by \citet{Li:2008ab,Li:2013aa}.

\subsection{Drift rate and duration}

The duration and drift rates of type III bursts are calculated from different aspects of the dynamic spectrum.  It is not immediately obvious that they would be related other than their dependence on frequency.  The drift rate decreases as a function of frequency, as shown by Figure \ref{figure:drift_rate}.  Similarly the duration increases as a function of frequency, as shown by Figure \ref{figure:duration}.  We can therefore assume that the drift rate will decrease as the duration increases.

\begin{figure}\center
\includegraphics[width=\wfig,trim=70 0 0 0,clip]{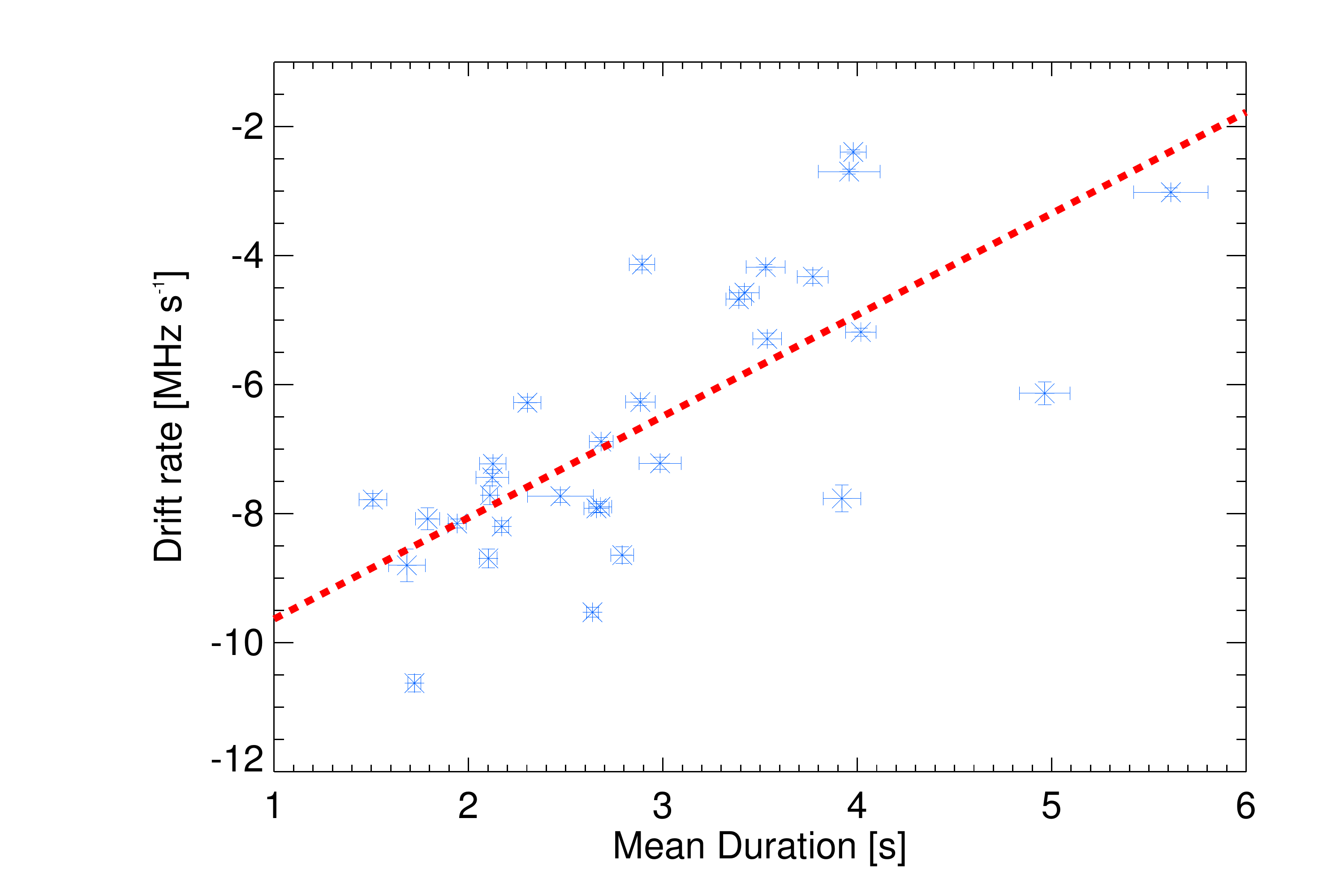}
\caption{Drift rate as a function of duration for each type III burst.  The red dashed line shows the linear fit to the data.}
\label{figure:drift_rate_duration}
\end{figure}

Characterising each individual type III bursts over the entire spectral range of 70 -- 30~MHz we can remove the dependence of frequency.  Figure \ref{figure:drift_rate_duration} compares the mean duration with the mean drift rate, found from the gradient of a linear fit to frequency with time.  There is a trend that radio bursts with faster drift rates have a smaller duration.  The correlation coefficient is 0.73. 

\section{Electron beam velocities} \label{sec:dynamics}

\subsection{Beam velocities}

To explain the temporal behaviour of the type III emission as a function of frequency we must consider the dynamics of the electron beam exciter and how it interacts with the background electron density to produce the Langmuir waves that are ultimately responsible for the observed radio waves.  The type III drift rate is related to the background electron density gradient $\pdv{n_e}{x}$ and the electron beam velocity $\pdv{x}{t}$ such that:
\begin{equation}\label{eqn:drift_rate}
\pdv{f}{t} = \pdv{f}{n} \pdv{n_e}{x} \pdv{x}{t}
\end{equation}
where $x$ is the radial propagation distance and $n_e$ is the background electron density that produces radio frequency $f$.  From Equation \ref{eqn:drift_rate} we can see that to obtain frequency drift rates with a smaller magnitude than \citep{Alvarez:1973ab}, we are observing either beams with smaller velocities or flux tubes with shallower density gradients.

% Assuming that any scattering effects of the radio emission is not significant to influence the rise time, the distribution of the electron beam is likely to play the dominant role.  The background density gradient will not likely be changing on timescales comparable to the electron beam velocity.  If we consider an electron beam that is extended in space over some finite length, the rise time is dependent upon the duration between when the front of the electron beam starts to produce significant radio emission at a certain frequency and when the middle of the electron beam produces the peak flux of radio emission.  If the rise time increases as a function of decreasing frequency, this time period must increase with radial distance from the Sun.  The velocity of the front of the electron beam must therefore be faster than the velocity of the middle of the electron beam.

% If the distribution of electrons is the main source of the decay time then a similar effect will occur at the back of the beam.  As the decay time increases as a function of decreasing frequency, the velocity at the back of the beam must therefore be slower than the velocity at the middle of the beam.  However, a number of other physical effects could also play a role in determining the decay time.  The decay time could be influenced by the scattering of light from source to observer.  Moreover, the decay time could be related to Langmuir waves not being re-absorbed by the electron beam and lingering in space after the beam has moved on.

We can estimate the beam velocities with a straight line fit to distance and time using Equation \ref{eqn:drift_rate} if we assume the \citet{Parker:1958aa} density model using a normalisation coefficient given by \citet{Mann:1999aa} and second harmonic emission.  The same density profile for all radio bursts introduces a non-statistical error into any beam velocities; the background density gradient will not likely be changing on timescales comparable to the electron beam velocity but will be different from burst to burst.  The peak time $t_0$ will provide an estimate of the velocity at the middle of the electron beam.  The leading edge of the electron beam will cause the generation of the first radio waves in time.  Consequently, a velocity for the front of the electron beam can be found by using the peak time minus the rise time, $t_0-t_{\rm r}$.  Similarly a velocity for the back of the beam can be found from the peak time plus the decay time, $t_0+t_{\rm d}$.

\vspace{20pt}
\begin{center}
\begin{table}
\centering
\caption{Velocities for the front, middle and back of the electron beams using the Parker density model.}
\begin{tabular}{ c  c  c}

\hline\hline

Velocity [c] 	& Mean 			& Standard Deviation \\ \hline
% rise 	& $0.008^{+0.005}_{-0.003}$		& $1.75\pm0.13$ \\
% peak 	& $0.01^{+0.007}_{-0.004}$		& $1.67\pm0.14$ \\
% decay 	& $0.005^{+0.003}_{-0.002}$		& $1.82\pm0.12$ \\
front 			& $0.20$		& $0.06$ \\
middle 			& $0.17$		& $0.05$ \\
back 			& $0.15$		& $0.04$ \\

\hline
\end{tabular}
\vspace{20pt}
\label{tab:velocity}
\end{table}
\end{center}

\begin{figure}\center
\includegraphics[width=\wfig,trim=100 0 0 0,clip]{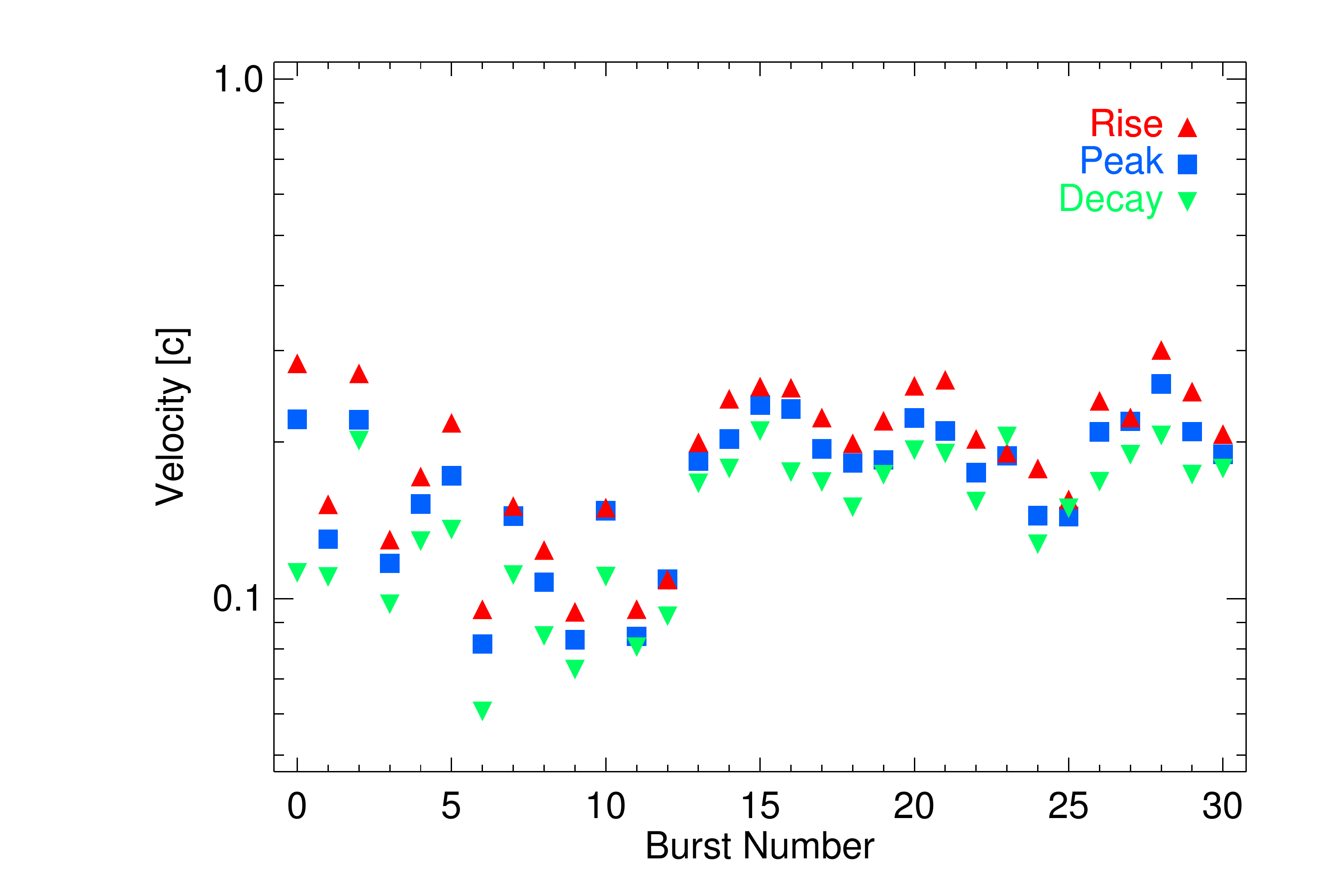}
\caption{Velocities derived from the rise, peak and decay times of each type III burst assuming a constant velocity and the Parker density model.  For almost all the bursts, the rise time is faster than the peak velocity which in turn is faster than the decay velocity.  The errors on the velocities from the fit are smaller than the symbols.}
\label{figure:rpd_velocities}
\end{figure}

Calculating the velocities of the electron beams from the rise, peak and decay times, Figure \ref{figure:rpd_velocities} shows the velocities for each type III burst using the Parker model.  The mean and standard deviation for the velocities are shown in Table \ref{tab:velocity}.  The common trend is that the rise velocity is faster than the peak velocity which is faster than the decay velocity.  This is expected behaviour as the bump-in-tail instability that drives the production of Langmuir waves is caused by faster electrons out pacing the slower electrons.  The rise time is dependent upon the duration between when the front and middle of the electron beam produce significant radio emission at a certain frequency.  If the rise time increases as a function of decreasing frequency, this time period increases with radial distance from the Sun.  The velocity of the front of the electron beam will therefore be faster than the velocity of the middle of the electron beam.  A similar effect occurs between the middle and back of the electron beam.

Our velocity calculations are assuming a 1--1 relation between the intensity of Langmuir waves and the intensity of radio emission.  Numerical simulations of type III bursts from an initial Maxwellian beam \citep[e.g.][]{Li:2011aa} and power-law electron beams \citep{Li:2011aa,Ratcliffe:2014aa} show close similarities between the speed of the electrons and the inferred speed of the exciter from synthetic radio emission.  Langmuir waves $L$ excited by faster electrons at the front of the beam are attributed to the onset of fundamental radio emission in \citep{Li:2013aa} due to the process $L\rightarrow T+S$ for transverse waves $T$ and ion-sound waves $S$.  However, there are differences.  For example, the production of second harmonic emission is intrinsically linked to the number of back-scattered Langmuir waves $L'$ able to take part in the $L+L'\rightarrow T$ process \citep{Ratcliffe:2014aa}.  Back-scattered Langmuir waves can be produced via Langmuir wave decay involving ion-sound waves by the processes $L+S\rightarrow L'$ and $L\rightarrow L'+S$.

The velocity calculations assume that scattering effects of the radio emission from source to observer do not significantly influence the rise time or the decay time.  When scattering is important, scattering is likely to broaden the time profile \citep{Steinberg:1971aa}, decreasing the measured drift rate of type III bursts and slow down any inferred velocities.  The effect on the rise and decay drift rates might be different; quantifying these effects are beyond the scope of this work.

The decay time could also be related to Langmuir waves not being re-absorbed by the electron beam and lingering in space after the beam has moved on.  In this instance the background density inhomogeneity will be the primary process in determining the exponential decay rate \citep[e.g.][]{Ratcliffe:2014aa}.

\subsection{Velocity and duration}

\begin{figure}\center
\includegraphics[width=\wfig,trim=70 0 0 0,clip]{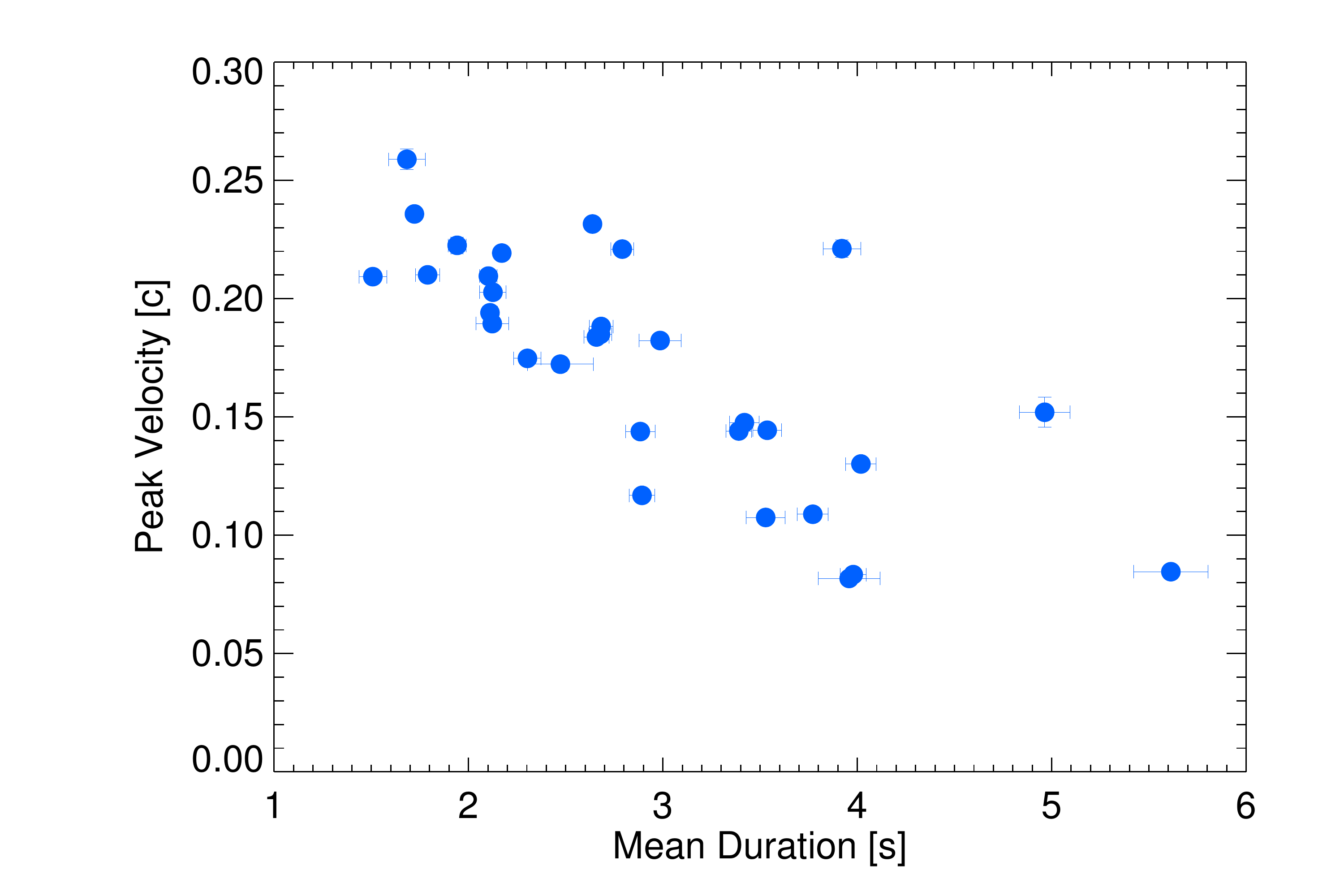}
\caption{Peak velocities as a function of durations for each type III burst.}
\label{figure:vel_dur}
\end{figure}

The velocity of the type III burst is directly related to the average drift rate.  Given the observational correlation between drift rate and duration in Figure \ref{figure:drift_rate_duration}, we show the peak velocity as a function of the duration for each type III burst in Figure \ref{figure:vel_dur}.  Both variables are correlated with a coefficient of 0.73, giving the appearance that faster electron beam create radio emission with a shorter duration.  The correlation is preserved if we use the front or back velocities, with the latter having a slightly higher correlation coefficient.

The simplest explanation for this correlation is that when the type III duration is smaller, higher energy electrons are driving the radio emission within the FWHM of the type III burst.  The corresponding duration the electrons spend at any one point in space will therefore be lower.  We discuss the type III duration and electron beam transport in more depth in Section \ref{sec:duration}.

\subsection{Beam length velocities}

If the front of the electron beam travels faster than the back of the electron beam then over time the electron beam will be stretched radially in space to cover a longer length.  The difference in electron beam velocity derived from the rise and decay times, $\Delta v=v_r-v_d$, will drive this elongation such that the beam will expand at a rate $t\Delta v$.  This assumes that all velocities are constant in time. 

\begin{figure}\center
\includegraphics[width=\wfig,trim=50 0 0 0,clip]{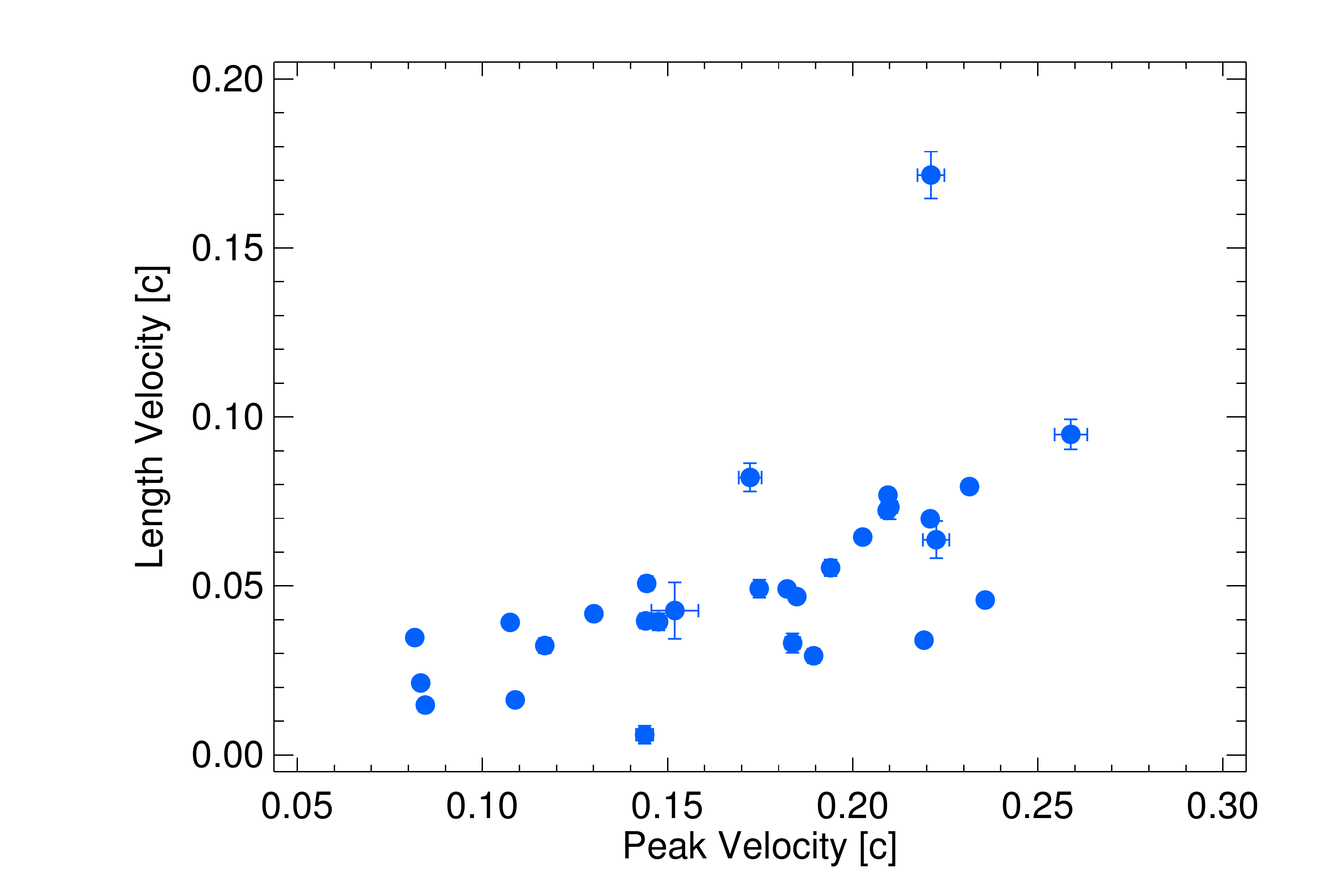}
\caption{Length velocity (rate at which the electron beam length increases) against the peak velocity.  The beam length velocity is derived from the difference between the velocities found from the rise and decay times.}
\label{figure:vel_dvel}
\end{figure}

In Figure \ref{figure:vel_dvel} we have plotted the length velocity $\Delta v$ against the velocity found from the peak.  We observe a linear correlation such that faster electron beams have a larger spread between the front and back velocities.  The data point in Figure \ref{figure:vel_dvel} with a high length velocity is caused by a bright type III burst that had very large decay times between 35--45 MHz.  This caused the resulting decay velocity to be anomalously small and consequently the length velocity was particularly large.  The Pearson correlation coefficient is 0.62 or 0.72 excluding the burst with anomalously high length velocity.

The origin of the correlation is related to the beam-plasma structure \citep[e.g.][]{Melnik:1998aa} that is formed by a propagating electron beam.  If the peak velocity is faster then higher energy electrons are involved in wave-particle interactions with the Langmuir waves which move faster through the background plasma.  Consequently faster velocities of electrons will be involved in the wave-particle interactions at the front of the electron beam.  However, the velocity at the back of the electron beam is more dependent upon the background thermal velocity, where Landau damping will ultimately absorb the Langmuir waves with low phase velocities.  The difference between the front and back velocities would therefore be larger for beams with higher peak velocities.

The length of the electron beam is directly related to the frequency bandwidth.  Therefore there is a connection between how the frequency bandwidth changes as a function of peak frequency and the length velocity.  We can represent the frequency bandwidth using $\Delta f = f_r - f_d$, where $f_r,f_p, f_d$ are the rise, peak and decay frequencies at any point in time.  The ratio of the bandwidth drift to the frequency drift is then
\begin{equation}
\frac{\pdv{\Delta f}{t}}{\pdv{f_p}{t}} = \frac{\pdv{f_r}{t}-\pdv{f_d}{t}}{\pdv{f_p}{t}}.
\end{equation}
Using Equation \ref{eqn:drift_rate} for the drift rate we can expand the ratio of bandwidth drift to drift rate to be
\begin{equation}
\frac{\pdv{f_r}{t}-\pdv{f_d}{t}}{\pdv{f_p}{t}} = \frac{\pdv{f_r}{n_r}\pdv{n_r}{x_r}\pdv{x_r}{t}-\pdv{f_d}{n_d}\pdv{n_d}{x_d}\pdv{x_d}{t}}{\pdv{f_p}{n_p}\pdv{n_p}{x_p}\pdv{x_p}{t}}.
\end{equation}
If we consider the bandwidth over a time range $\Delta t=t_1-t_0$ that is sufficiently large then the range of frequencies for the rise, peak and decay will be similar.  That is $f_r(t_0)\rightarrow f_r(t_1) \approx f_d(t_0)\rightarrow f_d(t_1)$.  In this instance we can assume the density gradients, $\partial n_r/\partial x_r$, $\partial n_p/\partial x_p$, $\partial n_d/\partial x_d$ are the same and that the derivatives $\partial f_r/\partial n_r$, $\partial f_p/\partial n_p$, $\partial f_d/\partial n_d$ are the same.  If we assume that the velocities $v=\partial x/\partial t$ are constant then the ratio of the bandwidth drift to the frequency drift can be expressed as
\begin{equation}
\frac{\pdv{\Delta f}{t}}{\pdv{f}{t}} = \frac{\pdv{f_r}{t}-\pdv{f_d}{t}}{\pdv{f}{t}} = \frac{\pdv{x_r}{t}-\pdv{x_d}{t}}{\pdv{x_p}{t}} = \frac{v_r-v_d}{v_p} = \frac{\Delta v}{v_p}.
\end{equation}
The ratio of the bandwidth drift to the peak frequency drift is therefore equivalent to the ratio of the length velocity to the peak velocity, independent of the density gradient.

\subsection{Non-constant velocities}

In all the above analysis, we assumed that the electron beam velocity remains constant over a finite distance of propagation.  Deceleration of the exciter has been inferred from the frequency and time of type III observations below 1~MHz \citep[e.g.][]{Fainberg:1972aa}, suggesting a deceleration of $12.3\pm0.8~\rm{km~s}^{-2}$ \citep{Krupar:2015aa}.

\begin{figure*}\center
\includegraphics[width=0.49\textwidth,trim=50 0 0 0,clip]{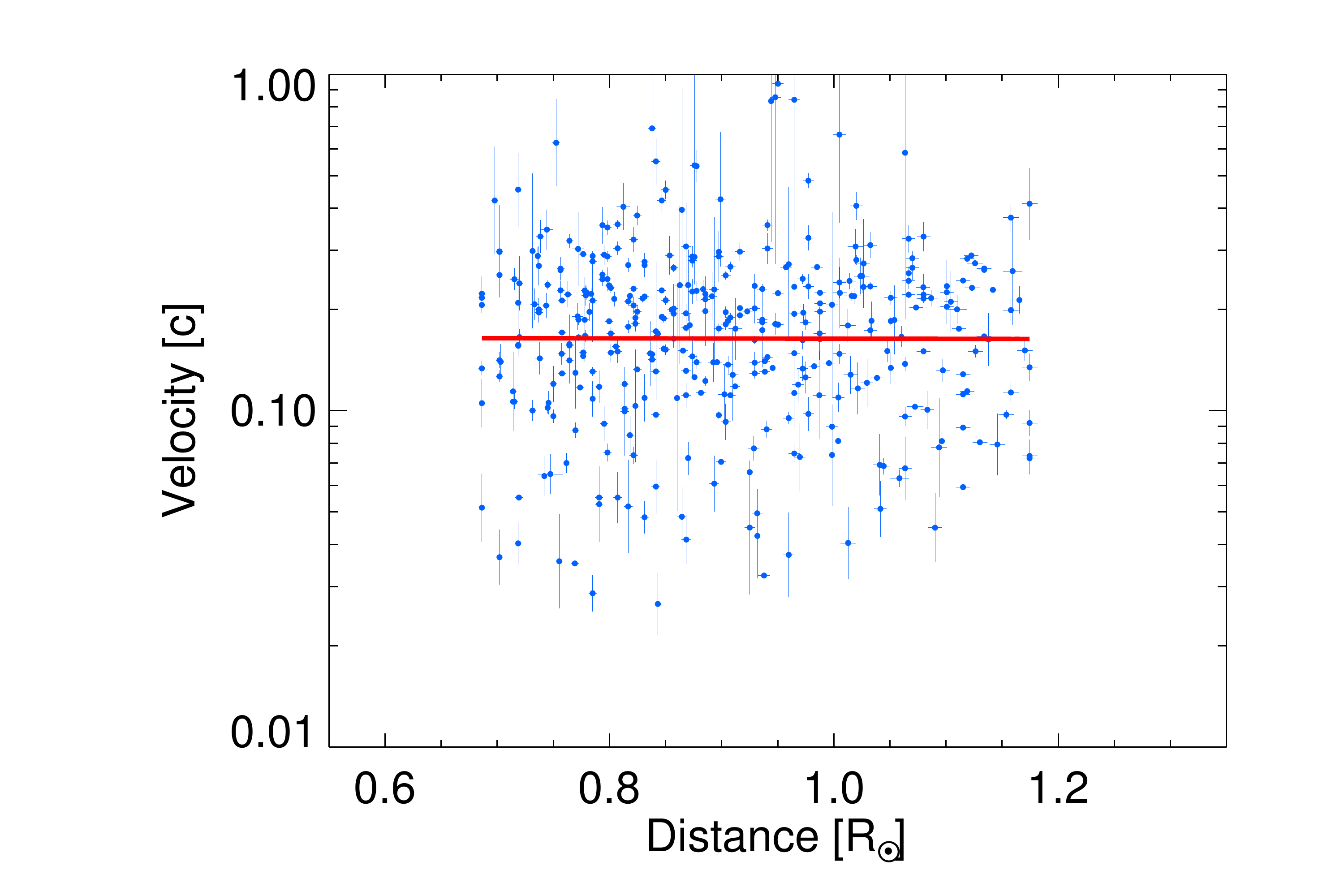}
\includegraphics[width=0.49\textwidth,trim=50 0 0 0,clip]{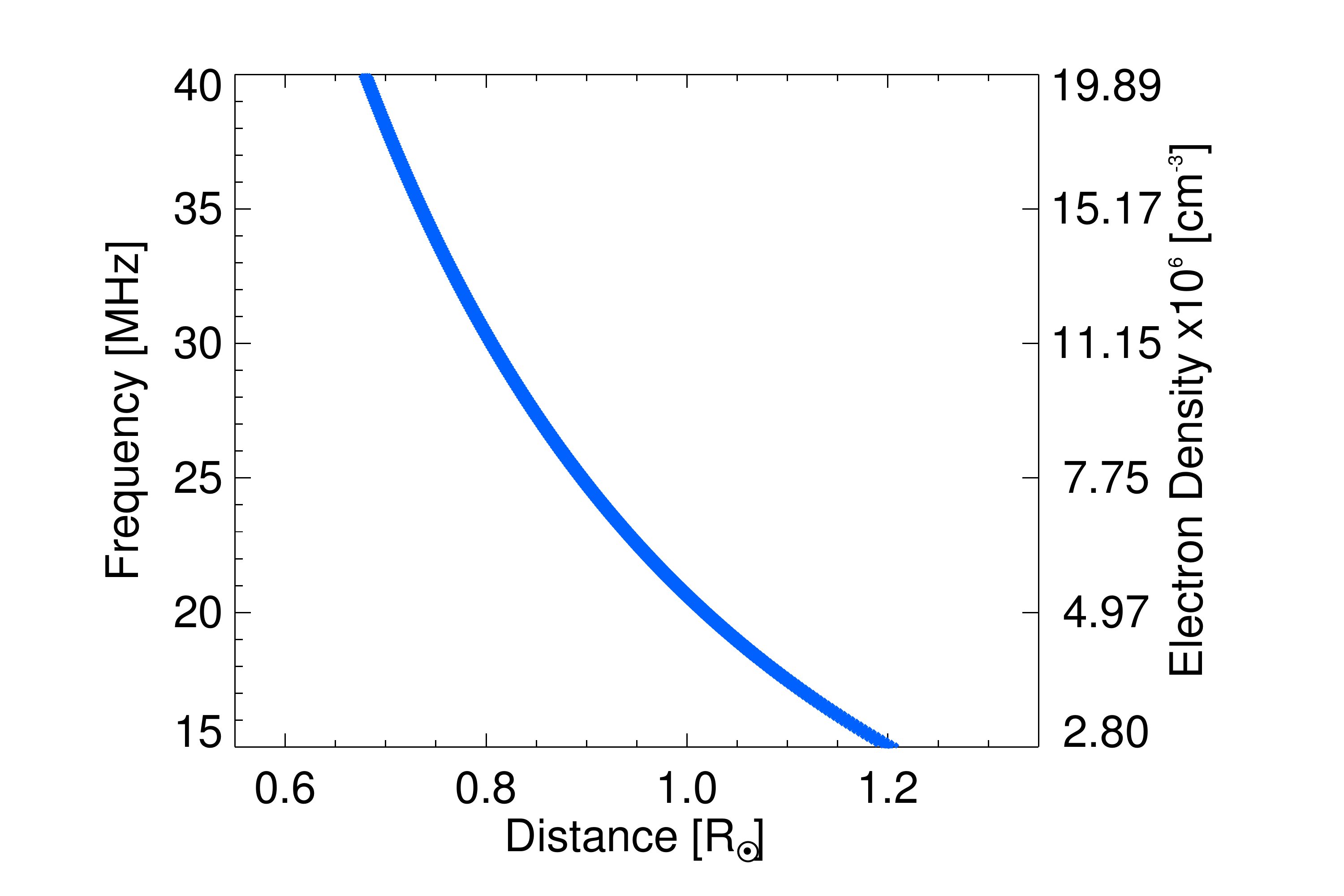}
\includegraphics[width=0.49\textwidth,trim=50 0 0 0,clip]{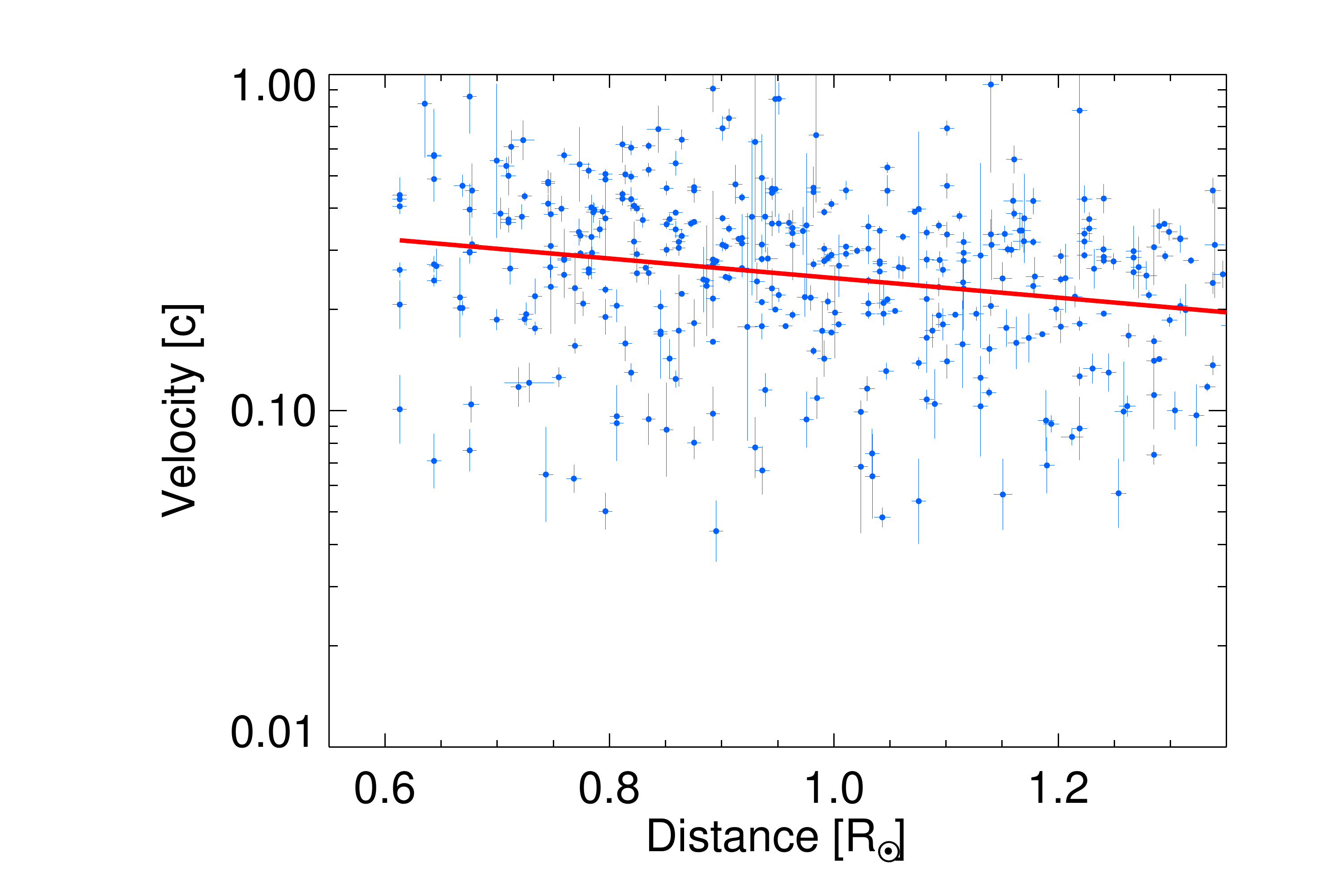}
\includegraphics[width=0.49\textwidth,trim=50 0 0 0,clip]{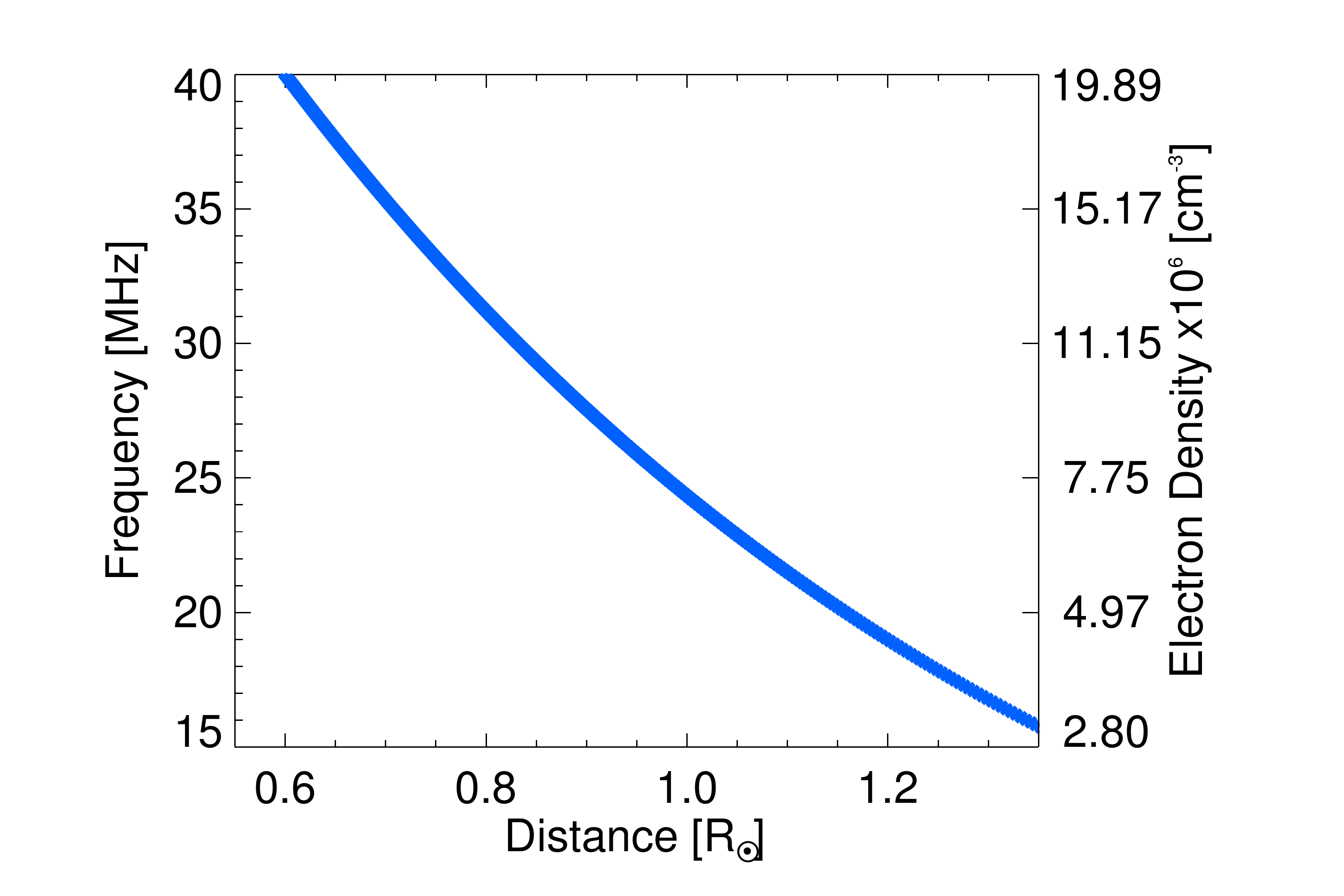}
\caption{Left: Velocities as a function of distance from the Sun for all the type III bursts using the Parker model (top) and the Dulk altitudes (bottom).  The red line is a linear fit to the log of the velocity such that $\log_{10}(v)=ar+c$.  Right: frequency and density profiles as a function of altitude for the Parker model (top) and Dulk model (bottom).}
\label{figure:drift_velocities}
\end{figure*}

To investigate whether or not there was a general trend for electron beam deceleration derived from higher frequencies, we found the velocity associated with the drift rate for each 3~MHz interval, in a similar way as the drift rate was obtained, over the frequency range 70--30~MHz.  Figure \ref{figure:drift_velocities} shows the velocity associated with the drift rates for all the type III radio bursts using the \citet{Parker:1958aa} density model and assuming second harmonic emission.  A linear fit to the log of the peak velocities such that
\begin{equation}
\log_{10}(v)=(-0.004\pm0.1)r-0.78\pm0.09.
\end{equation}
The change in velocity as a function of distance is very small, with the error on the gradient being significantly higher than the gradient itself.  The data therefore does not indicate a strong increase or decrease in the velocity of the electron beam as a function of distance.

A derived velocity is heavily dependent on the density model that is chosen.  Taking heights that were derived from observations by \citet{Dulk:1980aa} at 80 and 43 MHz of 0.6~$R_\odot$ and 1.2~$R_\odot$, respectively, we can construct a density model using the assumption that the density is exponentially decreasing with altitude.  The Dulk density model has a smaller density gradient than the model by \citet{Parker:1958aa}.  Consequently, the derived velocities are larger than those found previously.  The mean velocities of the type III bursts shown in Table \ref{tab:velocity} are approximately 0.08~c faster using the Dulk model, with similar standard deviations.

The second set of derived velocities is shown in Figure \ref{figure:drift_velocities} using the Dulk density model.  A linear fit to all the velocities shows a trend for the velocity of the exciter to decrease as a function of distance from the Sun.  Given the significant differences between the two different models, imaging spectroscopy is likely required to answer this question, taking into account the scattering effects of light from source to observer.  The analysis of fine frequency structures suggests significant size increases and a shift of the position radially \citep{Kontar:2017ab}.

% IDL> print,vel_fit_gau
%      -0.29050432    -0.024440810
% IDL> print,vel_fit_gau_err
% % PRINT: Variable is undefined: VEL_FIT_GAU_ERR.
% % Execution halted at: $MAIN$              1 /Users/hamish/lofar/L373207/expansion.pro
% IDL> print,r_fit_all_gau_err
%      0.063758785      0.12678062

\section{Electron beam transport} \label{sec:theory}

We have shown how the type III radio data can be used to estimate the velocities associated with the front, middle and back of the electron beam.  However, electron beams are not made up of mono-energetic electrons but are an ensemble of electrons spread over a range of energies.  Why can their travel be approximated by a single velocity and what exactly dictates how fast they travel?  How does this relate to the duration of the electron beam as a function of frequency?

\subsection{Gaussian Beam}

We consider the transport of a electron beam with a Gaussian distribution in velocity with mean velocity $v_0$ and characteristic velocity $\Delta v$, and a Gaussian profile in space with characteristic length $d$.  The initial electron distribution function is
\begin{equation}\label{eqn:source}
f(v,r,t=0) = n_{\rm beam} A_v \exp\left(-\frac{(v-v_0)^2}{2\Delta v^2}\right) \exp\left(-\frac{r^2}{d^2}\right),
\end{equation}
where $A_v=(\sqrt{2\pi}\Delta v)^{-1}$.  The mean velocity $v_0=10^{10}~\rm{cm~s}^{-1}$ corresponding to an energy of 28~keV, the density $n_{\rm beam}=10~\rm{cm}^{-3}$, the width $\Delta v=10^{9}~\rm{cm~s}^{-1}$ and the width $d=10^9~\rm{cm}$.
Concerning Langmuir wave generation, we present here the two limiting cases.  The first is the free-streaming approximation when the electrons propagate scatter-free.  The second is the gas-dynamic theory where the electrons fully interact with waves, propagating in a beam-plasma structure \citep[e.g.][]{Ryutov:1970aa,Kontar:1998aa,Kontar:2003aa}.

We consider only particle transport and the quasilinear interaction of particles and waves \citep{Vedenov:1963aa}.
\begin{equation}\label{eqn:quasi1}
\pdv{f}{t} + v\pdv{f}{r} = \frac{4\pi ^2e^2}{m_e^2}\pdv{}{v}\left(\frac{W}{v}\pdv{f}{v}\right),
\end{equation}
\begin{eqnarray}\label{eqn:quasi2}
\pdv{W}{t} = \frac{\pi \omega_{pe}}{n_e}v^2W\pdv{f}{v},
\end{eqnarray}
where the characteristic propagation time is $\tau_p=r/v$ and the characteristic quasilinear time is $\tau_{ql} = n_e/(\pi \omega_{pe} n_{\rm beam}) $.  The background electron density profile has the Parker density profile as in \citep{Kontar:2001ab,Reid:2010aa}.

\begin{figure*}\center

\includegraphics[width=0.49\textwidth]{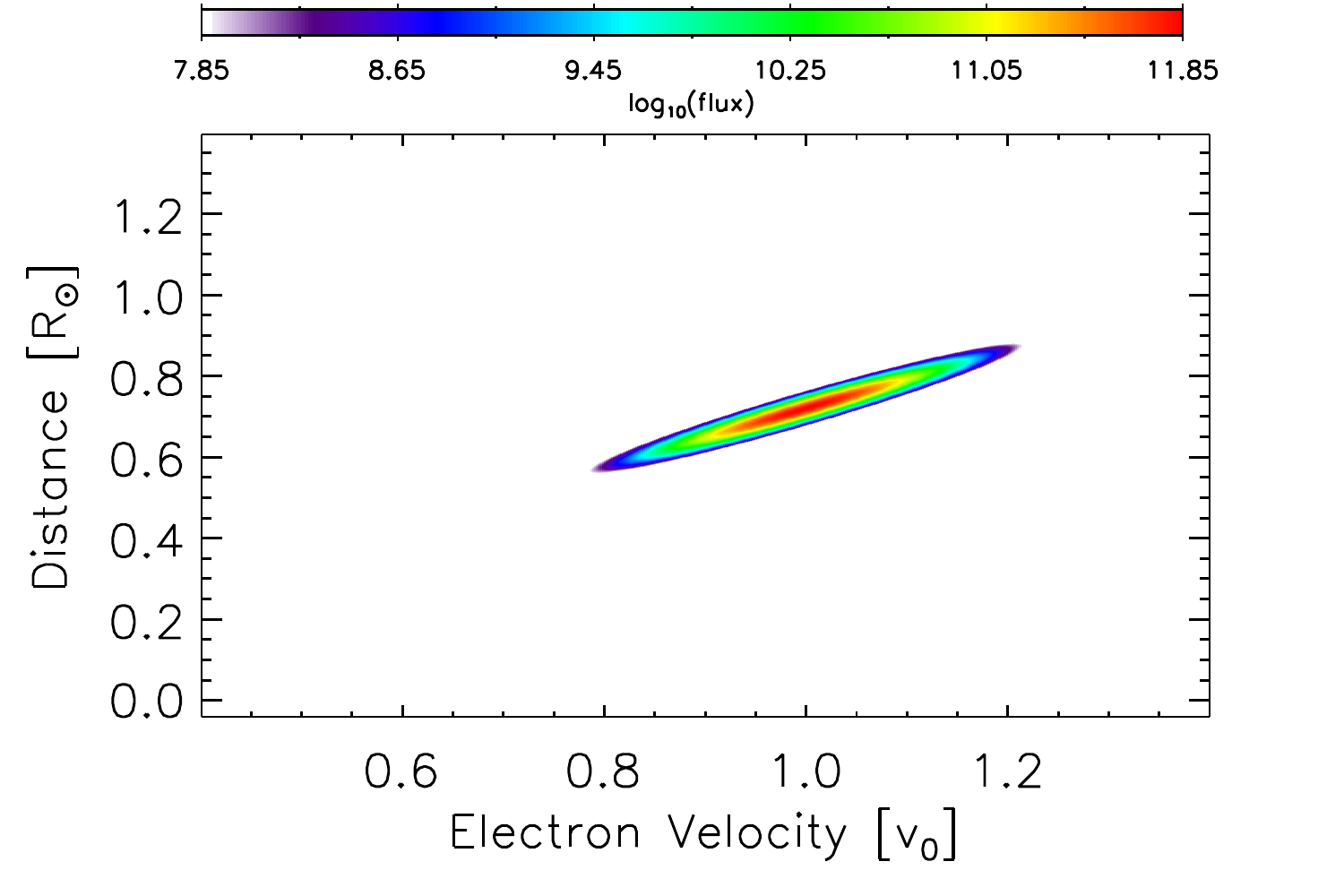}
\includegraphics[width=0.49\textwidth]{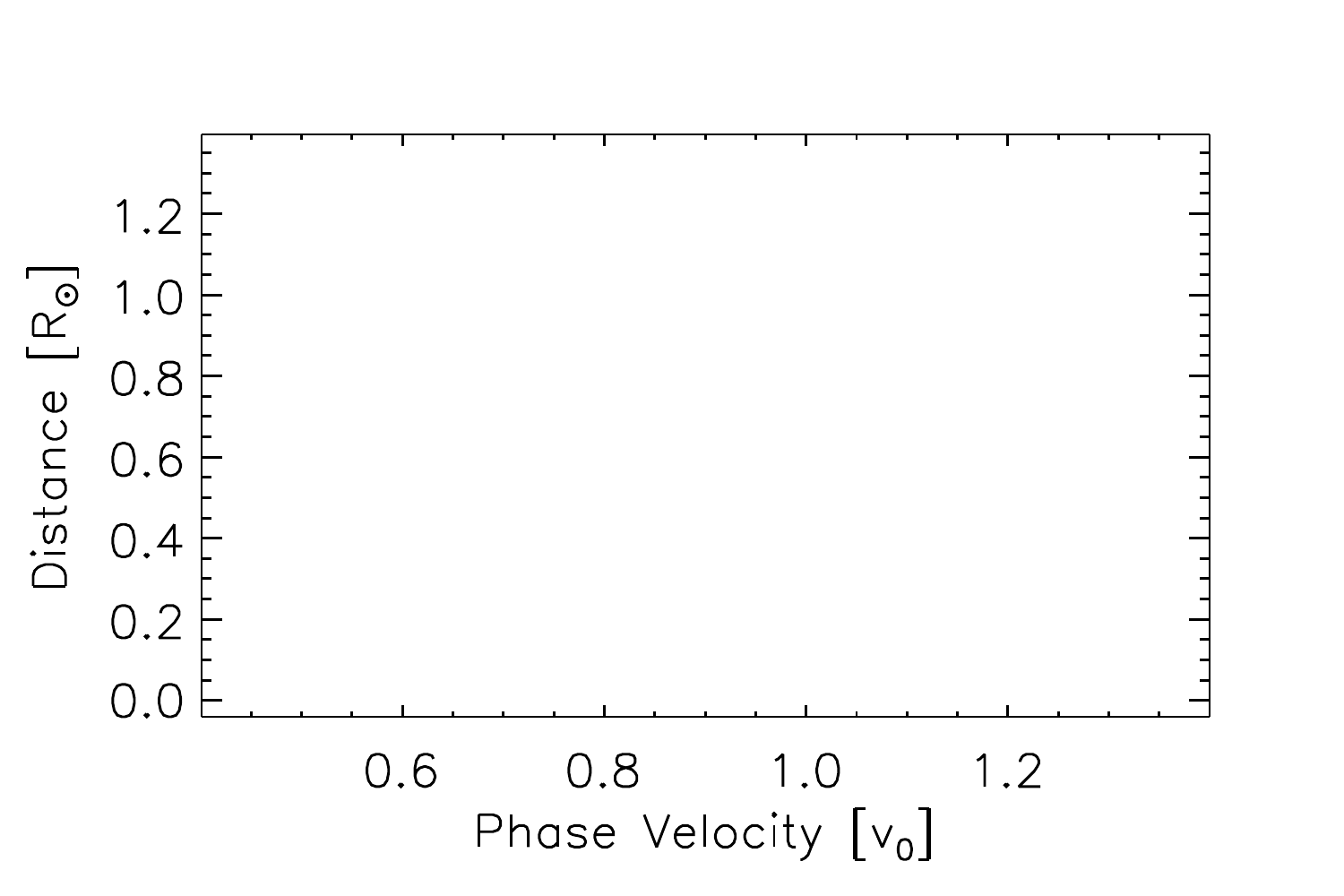}
\includegraphics[width=0.49\textwidth]{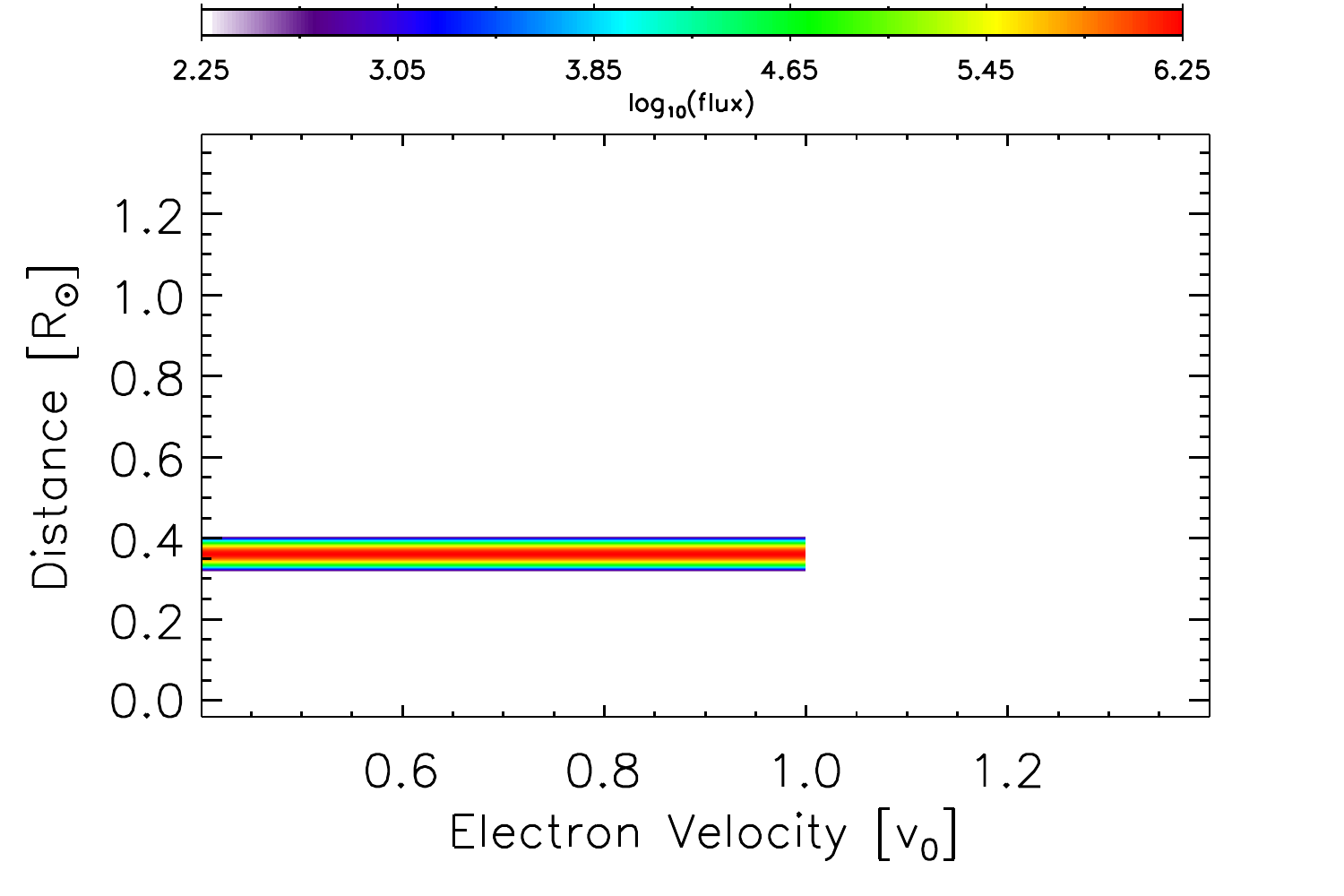}
\includegraphics[width=0.49\textwidth]{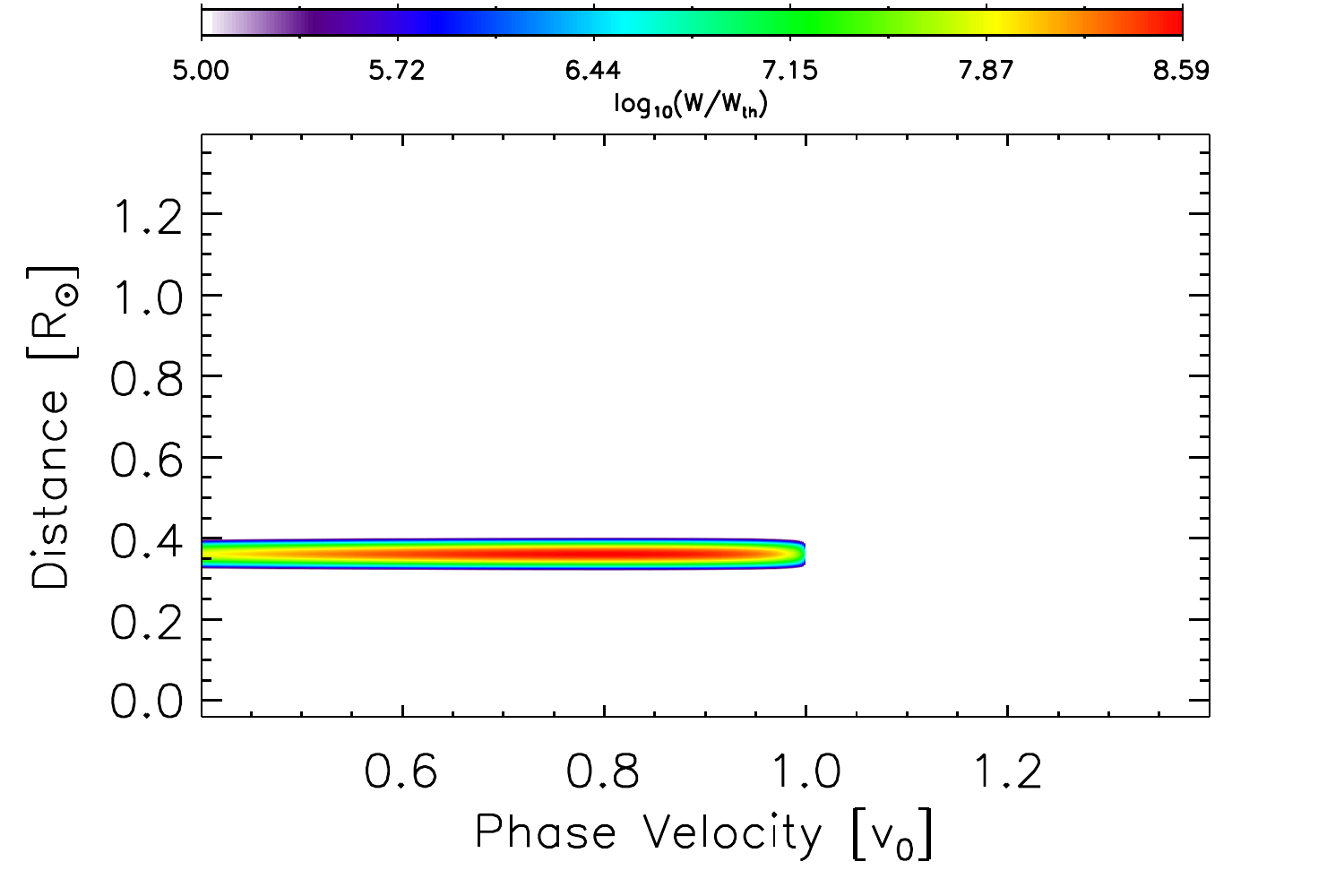}
\includegraphics[width=0.49\textwidth]{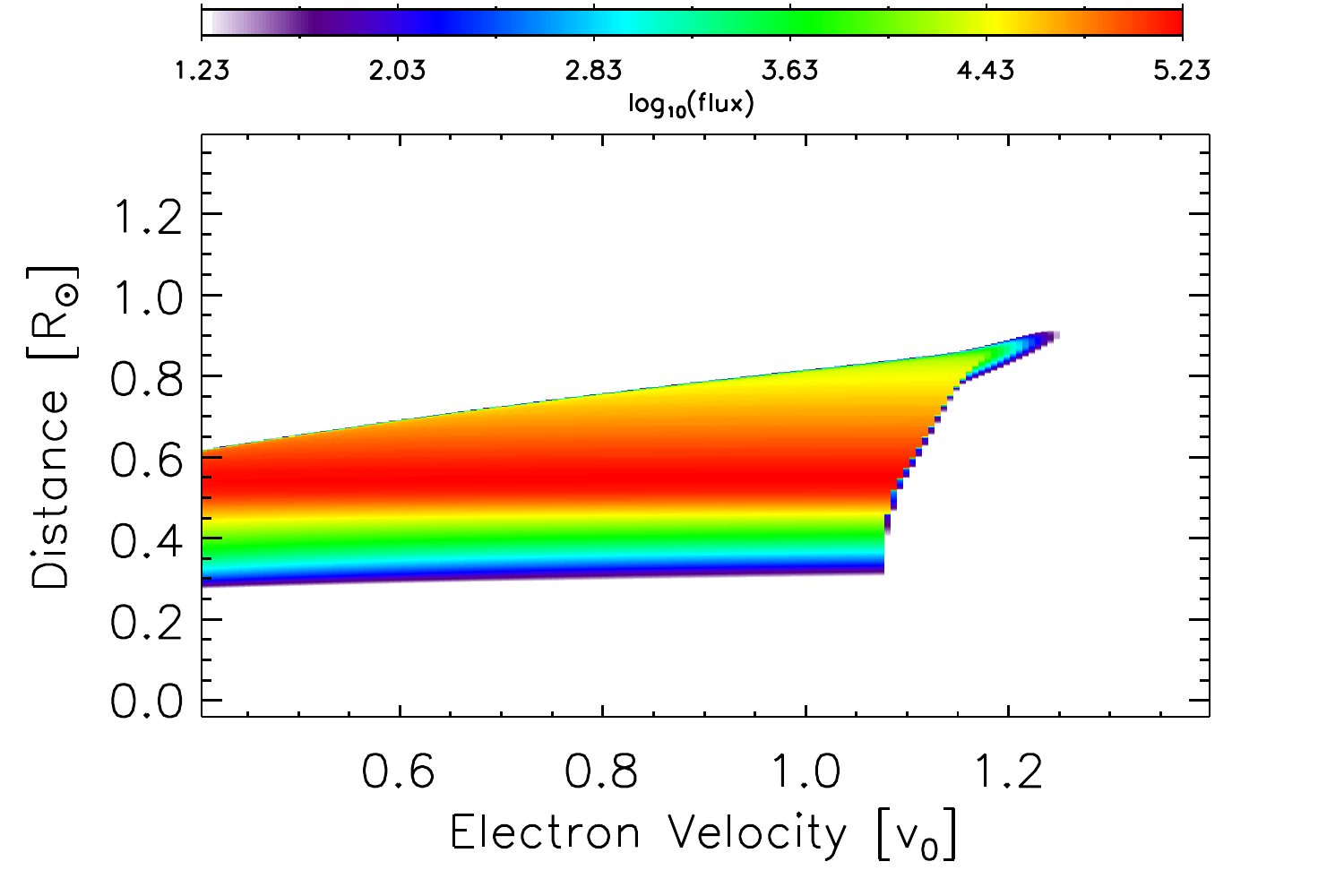}
\includegraphics[width=0.49\textwidth]{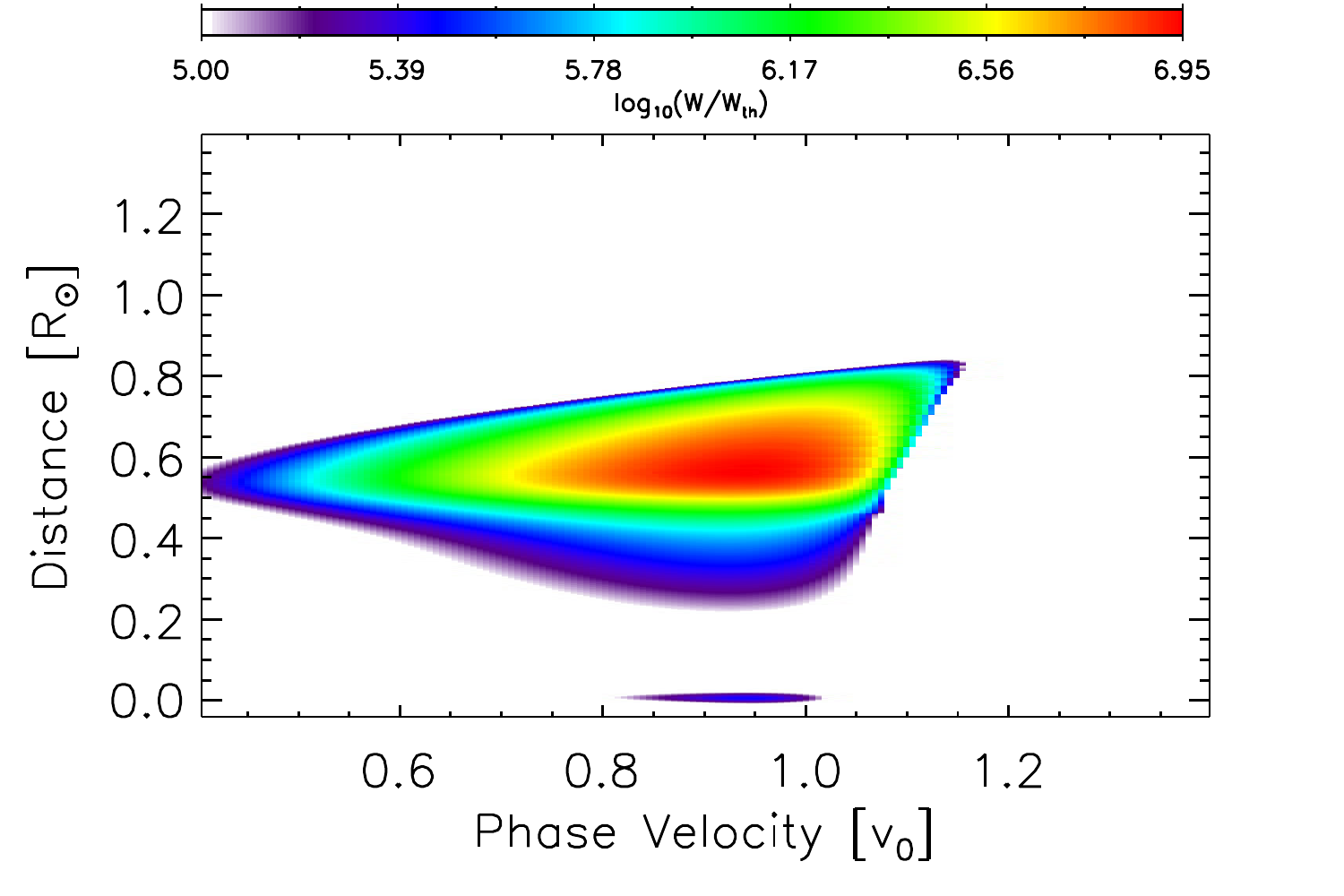}
\caption{Electron flux (left) and normalised spectral energy density (right) from an initial Gaussian electron beam with  $v_0=10^{10}~\rm{cm~s}^{-1}$ (0.33~c).  Top: free-streaming case with zero waves generated.  Middle: gas-dynamic theory with electron beam velocities $v<v_0$.  Bottom: simulations using Equations \ref{eqn:quasi1}, \ref{eqn:quasi2}.}
\label{figure:dist_gauss}
\end{figure*}

The first limiting case is when the electrons propagate without interacting with Langmuir waves.  This case essentially propagates with $\tau_{ql}=\infty$ so no waves are generated.  Figure \ref{figure:dist_gauss} shows the distribution function after 5 seconds.  The differences in velocity have elongated the electron beam in space.

The second limiting case is the gas-dynamic theory (GDT) that analytically describes the motion of a spatially limited electron beam propagating in a background plasma.  The GDT assumes that the quasilinear time $\tau_{ql}=0$.  A description of how a Gaussian beam evolves in space is given in \citet{Kontar:1998aa} for electrons with velocity $v<v_0$.  A plateau is produced with height $n_{\rm beam}/v_0$.  The distribution moves with a constant velocity of $(v_0+v_{\rm min})/2$.  A plateau in the electron distribution function is formed in velocity space at every point in position space.   The particles and waves propagate together in a beam-plasma structure at a constant velocity determined by the mean velocity in the plateau.  The structure can propagate over large distances as it conserves its spatial shape, energy and the number of particles.  Figure \ref{figure:dist_gauss} shows the distribution function after 5 seconds for velocities $v<v_0$.  There is no spread of the electron beam over distance as the collective motion of the electrons propagate at a single velocity.

The third case is a simulation of the electron beam that evolves using Equations \ref{eqn:quasi1} and \ref{eqn:quasi2}.  Figure \ref{figure:dist_gauss} shows the distribution after 5~seconds.  We can see that the simulation lies between the two limiting cases.  The highest velocities have mostly propagated scatter-free.  The lower velocities have formed a plateau and have generated Langmuir waves.  Elongation in space of the plateau occurs because the free-streaming electrons become more unstable to Langmuir waves during propagation.  The value of $\tau_{ql}$ varies in phase space and also depends upon a number of constants including $n_{\rm beam}$, $v_0$, $\Delta v$.  As an example, as $n_{\rm beam} \rightarrow 0$, the system will propagate as the free-streaming case.  As $n_{\rm beam} \rightarrow \infty$ the system will propagate in the gas-dynamic case.

\subsection{Power-law beam}

The signature of energetic electrons in the solar atmosphere deduced from X-rays infers a power-law injection function in velocity space \citep{Holman:2011aa}.  A complimentary power-law spectrum is measured in situ near the Earth \citep{Krucker:2007aa}.  As such, we model the injection of an electron beam with a power-law distribution in velocity space into a spatially limited acceleration region with a Gaussian profile.  The injection function can be described as
\begin{equation}\label{eqn:source2}
f(v,r,t=0) = A_v v^{-\alpha}\exp\left(-\frac{r^2}{d^2}\right),
\end{equation}
where $A_v=(\alpha-1)n_{\rm beam}/(v_{\rm min}^{(-\alpha+1)}-v_{\rm max}^{(-\alpha+1)})$ normalises the injection profile such that the number density injected at $r=0$ (acceleration region) is $n_{\rm beam}$ [$\rm{cm}^{-3}$].  The velocity distribution is characterised by $\alpha$, the spectral index, and the spatial distribution is characterised by $d$ [cm].

\begin{figure}\center
\includegraphics[width=\wfig]{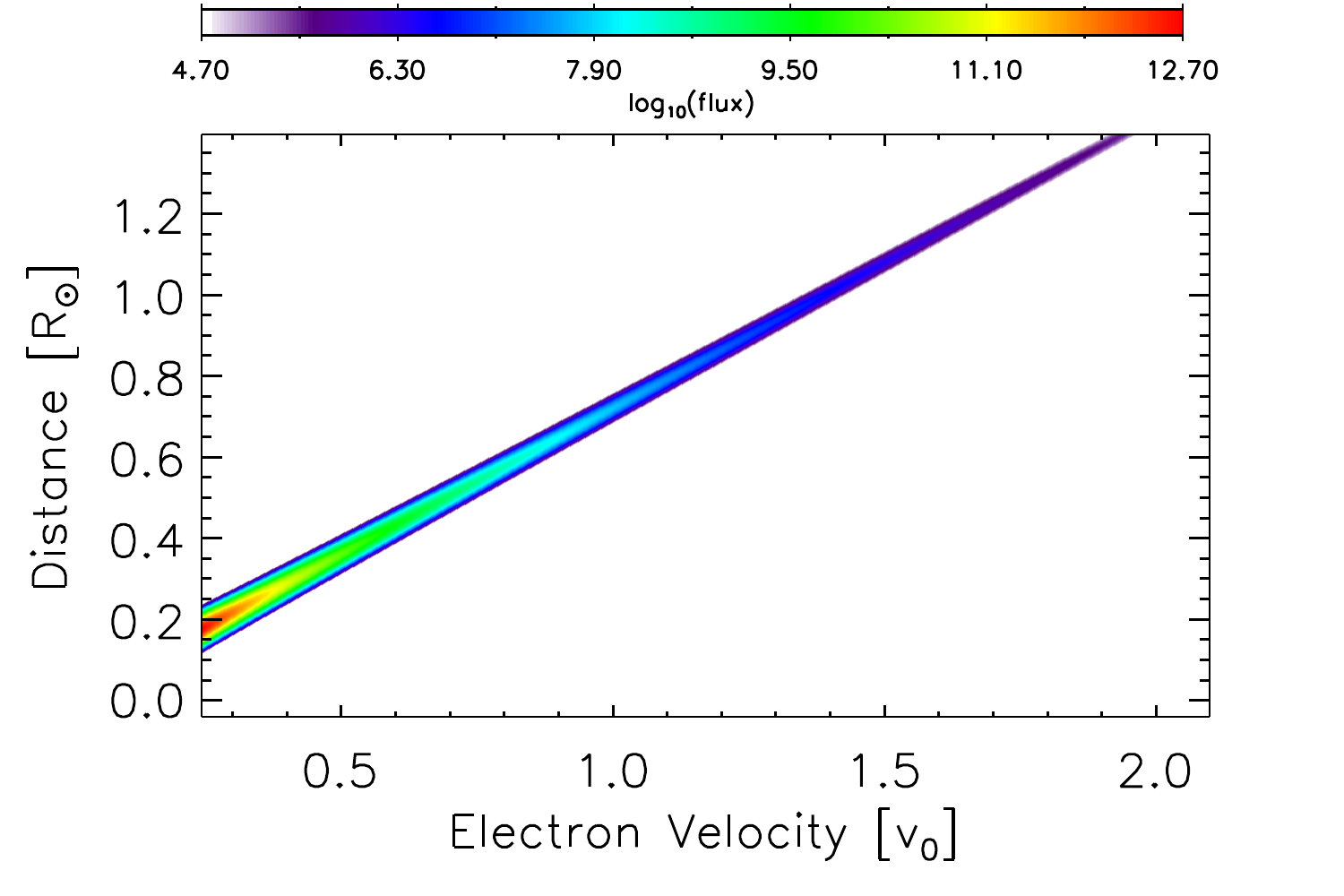}
\includegraphics[width=\wfig]{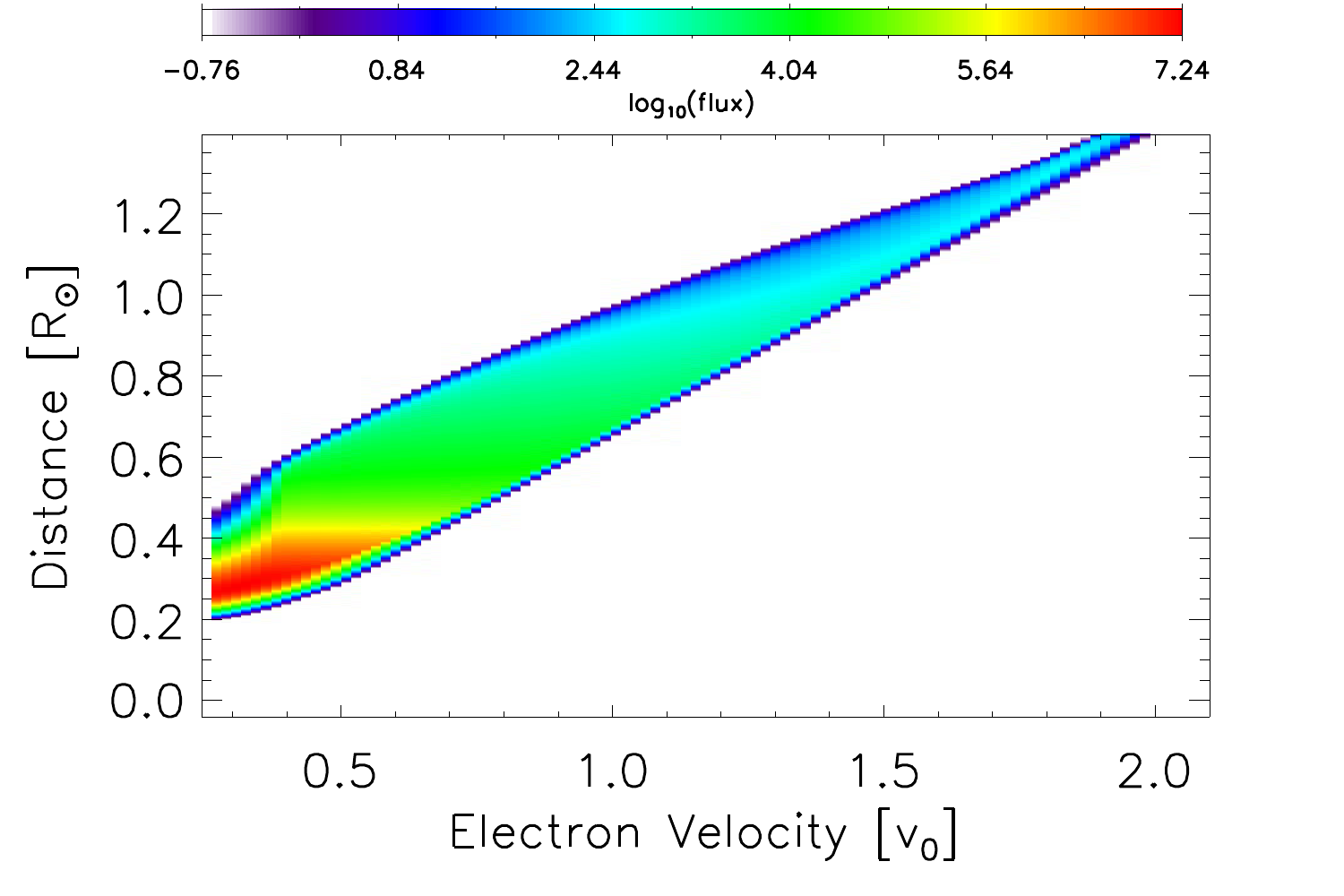}
\includegraphics[width=\wfig]{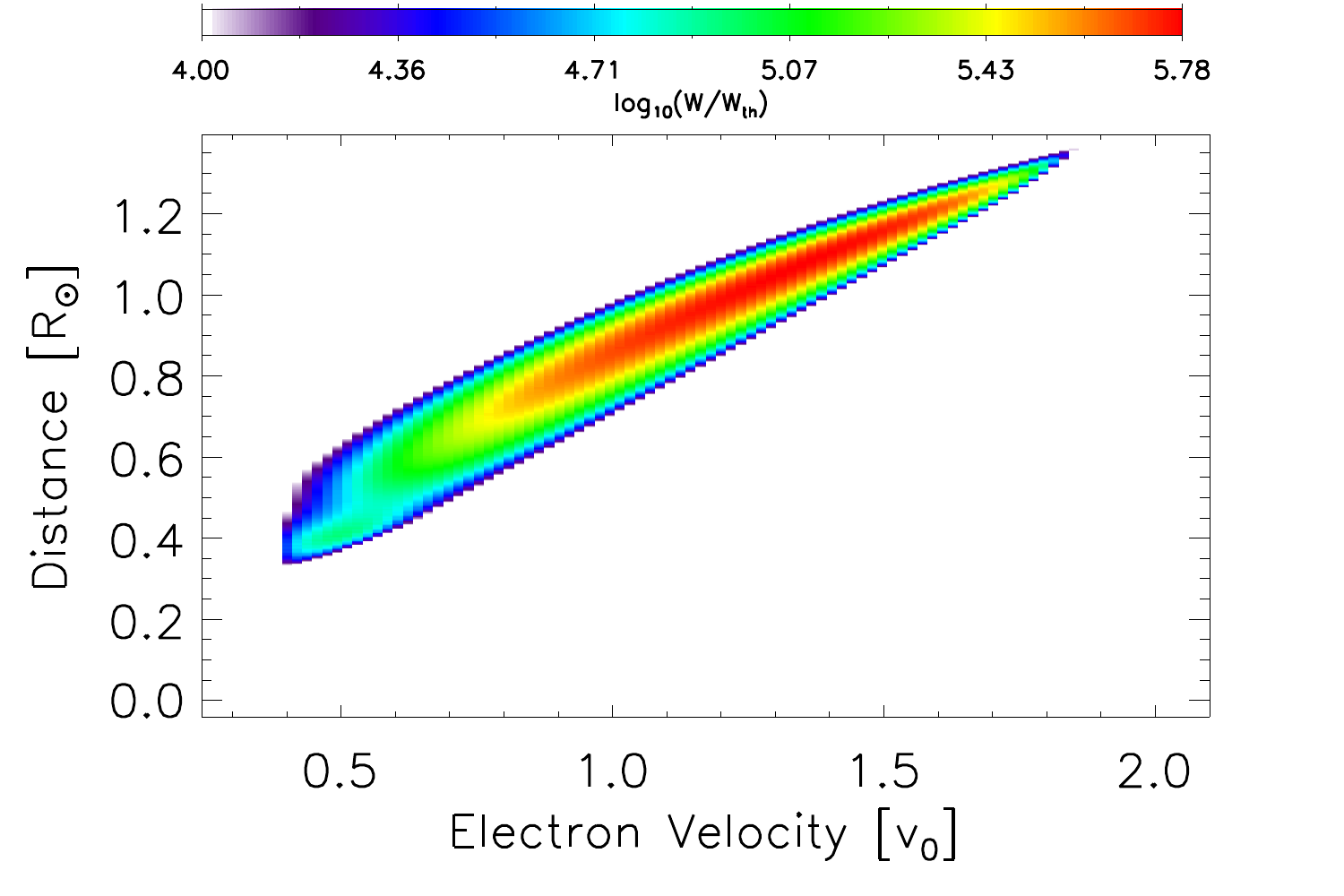}
\caption{Top: electron flux from a free-streaming electron beam with an initial power-law distribution in velocity after $t=5$~seconds ($v_0=10^{10}~\rm{cm~s}^{-1}$, 0.33~c).  Middle and bottom: simulated electron flux and spectral energy density from a propagating electron beam with an initial power-law distribution, with a set-up described in \citet{Reid:2015aa}, after $t=5$~seconds of propagation.}
\label{figure:distributions_sims}
\end{figure}

Using flare parameters of $\alpha=8$, $n_{\rm beam}=10^6~\rm{cm}^{-3}$, $d=10^9$~cm, $v_{\rm min}=2.4\times10^9~\rm{cm~s}^{-1}$ and $v_{\rm max}=2\times10^{10}~\rm{cm~s}^{-1}$ we have plotted a snapshot of the electron flux in Figure \ref{figure:distributions_sims} at time $t=5$~seconds, after free-streaming propagating, where $v_0=10^{10}~\rm{cm~s}^{-1}$.  It shows how the faster electrons outpace the slower electrons, stretching the electrons over a wide range of space.  Between two velocities $v_1,v_2$, where $v_1>v_2$, the width of the electron beam will increase in time and can be described by $(v_2-v_1)t+d$.  The duration of the electron beam at one point in space with thus increase as a function of distance from the Sun, as the background plasma frequency decreases.

We have previously simulated electron beams starting with an initial power-law injection of electrons and propagating out of the corona, interacting with Langmuir waves \citep{Kontar:2009aa,Reid:2010aa,Reid:2013aa,Reid:2015aa,Reid:2017ab}.  Using the same set-up as \citet[][Section 5]{Reid:2015aa} we show an electron beam that was injected with a beam density of $10^{6}~\rm{cm}^{-3}$.  Whilst the initial power-law injection function is stable, propagation effects cause Langmuir waves to be generated and the electron beam to form a plateau in velocity space.  Figure \ref{figure:distributions_sims} shows the electron flux and accompanying Langmuir wave spectral energy density after $t=5$~seconds.  The Langmuir waves spectrum has been normalised by the background thermal level.

The difference between the free-streaming and particle-wave electron distributions is clear from comparing the two electron flux plots in Figure \ref{figure:distributions_sims}.  In both scenarios the highest energy electrons are propagating scatter-free.  Below approximately $1.8~v_0$ the electrons are able to interact with Langmuir waves and the electron distribution function has relaxed to a plateau at many different points in space.  Each plateau has a different minimum and maximum velocity.

At the front of the beam, the largest velocity in the plateaus are at their highest.  The transport time (time that an electron beam spends at any one point in space) is sufficiently fast in comparison to the electron diffusion time in velocity space that the electron beam does not relax down to near thermal velocities.  The minimum velocity in the plateau is significantly higher than $v_{te}$.

As we progress down towards the back of the beam, the largest velocities in the plateaus get smaller.  Moreover, the diffusion of electrons in velocity space increases such that the smallest velocities become closer to $v_{te}$.  The resulting range of plateau velocities means that the front of the beam moves faster than the back of the beam.

The length of the electron beam will increase at a constant rate if the velocities within the plateau do not change.  Between plateaus with velocities $u_1=(u_{1,max}+u_{1,\rm{min}})/2$ and $u_2=(u_{2,max}+u_{2,\rm{min}})/2$, where $u_1>u_2$, the width will increase as $(u_1-u_2)t$ after wave-particle interactions become important.

\subsection{Type III durations} \label{sec:duration}

The duration derived from a free-streaming electron beam at one point in space is dependent upon the velocity of electrons that are considered.  Between velocities $v_1, v_2$, where $v_1>v_2$, at some distance $r$ from the injection site, the duration is simply
\begin{equation}
\rm{t}_D^{fs} = r/v_2-r/v_1.
\end{equation}
From the LOFAR radio data, we can estimate the velocities $v_1$ and $v_2$ from the rise and decay velocities, respectively using the mean rise velocity $v_1=0.20$c and the mean decay velocity $v_2=0.15$c (Table \ref{tab:velocity}).

\begin{figure}\center
\includegraphics[width=\wfig,trim=30 0 0 0,clip]{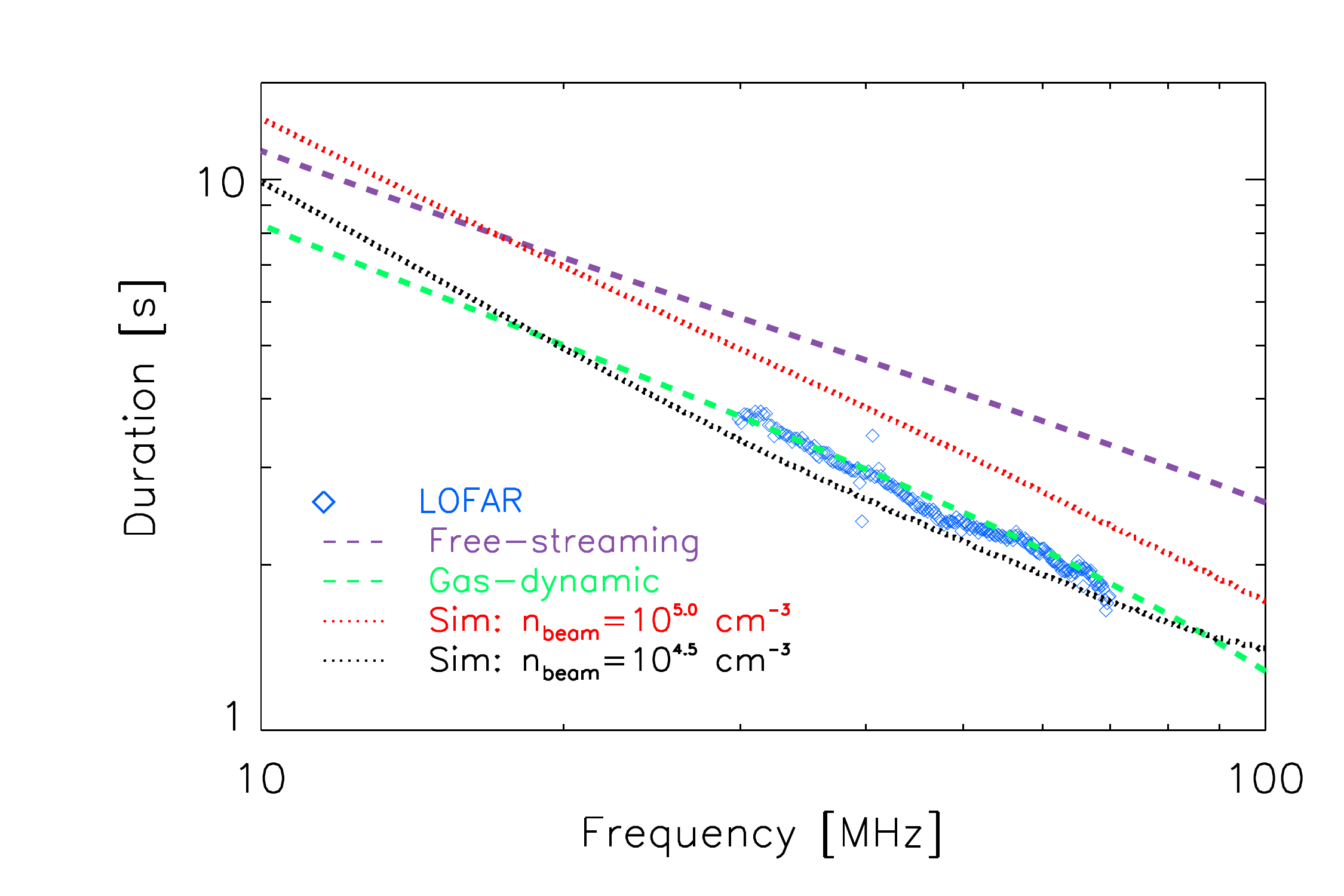}
\caption{Electron beam duration at different plasma frequencies, assuming second harmonic emission.  The free-streaming curve is calculated assuming $v_1=0.20$~c and $v_2=0.15$~c.  The Gas-dynamic fit is calculated assuming free-streaming to 100~MHz using $u_{1,max}=0.24$~c and $u_{2,max}=0.19$~c then plateaus with average velocities $u_1=0.20$~c and $u_2=0.15$~c.  The simulation with $n_{\rm beam}=10^{4.5}~\rm{cm}^{-3}$ (black) has $u_1=0.18$~c and $u_2=0.14$~c between 70--30~MHz.  The simulation with $n_{\rm beam}=10^{5.0}~\rm{cm}^{-3}$ (red) has $u_1=0.22$~c and $u_2=0.14$~c between 70--30~MHz.}
\label{figure:free_streaming}
\end{figure}

Figure \ref{figure:free_streaming} shows the corresponding free-streaming duration as a function of plasma frequency assuming the Parker density model, the emission is second harmonic, and that the electron beam starts at an altitude of $50$~Mm, an altitude found from fitting radio and X-ray data \citep{Reid:2014aa}.  The free-streaming duration using velocities $v_1$ and $v_2$ is incompatible with the mean radio burst durations, having a significantly higher duration at all frequencies.  This result is unsurprising given that the free-streaming approximation misses out the critical physical processes involved in radio burst production.

The duration derived assuming gas-dynamic theory will be constant as a function of frequency if we assume only one plateau with a constant velocity.  However, if we assume that the front and back of the beam moves with plateau velocities $u_1=(u_{1,\rm{max}}+u_{1,\rm{min}})/2 = 0.20$c and $u_2=(u_{2,\rm{max}}+u_{2,\rm{min}})/2 = 0.15$c, respectively, the duration will increase as a function of decreasing frequency.  An electron beam with a power-law distribution must propagate an instability distance before Langmuir waves are produced \citep{Reid:2011aa,Reid:2013aa,Reid:2014aa}.  The beam can be assumed to have propagated this instability distance under free-streaming conditions with velocities $u_{1,\rm{max}}$ and $u_{2,\rm{max}}$.  Velocity diffusion can then produce the plateaus $u_1$ and $u_2$ when Langmuir waves are generated.  Given an instability distance $r_c$, the corresponding duration of an electron beam can be found from
\begin{equation}\label{eqn:gdt_dur}
\rm{t}_D^{gd} = \left(\frac{r-r_c}{u_2}+\frac{r_c}{u_{2,\rm{max}}}\right) - \left(\frac{r-r_c}{u_1}+\frac{r_c}{u_{1,\rm{max}}}\right), \quad r\geq r_c.
\end{equation}
We assume that $r_c=4.13\times10^{10}$~cm, ($0.59~R_{\odot}$), which corresponds to a plasma frequency of 50~MHz in the Parker density model that would produce second harmonic radio emission at 100~MHz.  We can then fit the LOFAR durations using $u_{1,\rm{min}}=0.15$~c and $u_{2,\rm{min}}=0.10$~c which gives $u_{1,\rm{max}}=0.24$~c and $u_{2,\rm{max}}=0.2$~c.  The value of $u_{2,\rm{min}}=0.10$~c is approximately $5.5~v_{te}$ for a 2~MK plasma and $u_{1,\rm{min}}>u_{2,\rm{min}}$, as shown by the simulations in Figure \ref{figure:distributions_sims}.

We chose the value of $u_{1,\rm{min}}$ and $u_{2,\rm{min}}$ so that the corresponding values of $\rm{t}_D^{gd}$ overlap the LOFAR durations, shown in Figure \ref{figure:free_streaming}.  Taking into account higher velocities during the instability distance will always reduce the duration with respect to the free-streaming approximation, when $u_1=v_1$ and $u_2=v_2$.  However, it is important to note that our choice of $u_{1,\rm{min}}$ and $u_{2,\rm{min}}$ was arbitrary.  Assuming a higher/lower $u_{1,\rm{min}}$ will decrease/increase the value of $\rm{t}_D^{gd}$ accordingly for the same $u_1$, $u_2$.  Similarly increasing/decreasing $r_c$ with decrease/increase the value of $\rm{t}_D^{gd}$.  Despite $\rm{t}_D^{gd}$ over-plotting the radio data very well, the simple formula given by Equation \ref{eqn:gdt_dur} is only an approximation and does not take into consideration many different physical mechanisms.  It fits the radio data below 100 MHz well because it used the derived velocities of 0.20~c and 0.15~c estimated from the LOFAR data.

\vspace{20pt}
\begin{center}
\begin{table*}
\centering
\vspace{20pt}
\caption{Initial beam parameters for the electron beam injected into the solar corona.}
\begin{tabular}{ c  c  c  c  c  c}

\hline\hline

Energy Limits & Velocity Limits & Spectral Index & Temporal Profile & Spatial Profile & Expansion\\ \hline
1.7--113~keV &  $4-36~v_{\rm th}$ & $\alpha=10.0$& $\tau=1$~s & $d=10^9$~cm & $\beta=2.0$ \\

\hline
\end{tabular}
\label{tab:beam_sun}
\end{table*}
\end{center}

Numerical simulations of an electron beam can be used to estimate the corresponding durations at different frequencies.  We can use the spectral energy density of the Langmuir waves as a proxy for the radio emission as the intensity will be related to the intensity of radio emission.  For each frequency, we find the peak spectral energy density at each point in time.  The duration is then calculated from the corresponding FWHM.  Using the Langmuir waves as a proxy for radio emission ignores the effects of the wave-wave processes generating the radio waves from the Langmuir waves.  We must therefore treat any results derived from this with care.

We used simulations described in \citet{Kontar:2001ad,Reid:2013aa,Reid:2015aa}, with initial conditions described by Table \ref{tab:beam_sun}.  We simulated two different initial beam densities $n_{\rm beam}=10^{4.5}~\rm{cm}^{-3}$ and $n_{\rm beam}=10^{5.0}~\rm{cm}^{-3}$.  Varying the beam density changes the average velocities of the plateaus at the front of the beam.  Higher beam densities allow electrons with higher velocities to produce Langmuir waves at the front of the electron beam.  The velocity at the back of the electron beam is dominated by the thermal velocity.

The corresponding durations from the quasilinear simulations are shown in Figure \ref{figure:free_streaming}.  The simulation with $n_{\rm beam}=10^{4.5}~\rm{cm}^{-3}$ resulted in a front velocity of $u_1=0.18$~c and a back velocity of $u_2=0.14$~c between plasma frequencies 35--15~MHz, relating to 70--30~MHz assuming second harmonic emission.  These velocities are similar to the LOFAR observations.  There is a corresponding agreement between the durations derived from this simulation and the durations from the LOFAR observations.  We note that the initial beam density was chosen such that $u_1$ and $u_2$ would be similar to the LOFAR observations.  

The simulation with $n_{\rm beam}=10^{5.0}~\rm{cm}^{-3}$ resulted in a front velocity of $u_1=0.22$~c and a back velocity of $u_2=0.14$~c between the same plasma frequencies, faster than the LOFAR observations.  The corresponding derived durations from the Langmuir waves are longer than the LOFAR observations.  Whilst it is tempting to infer that higher beam densities result in increased durations, we note that the type III observations inferred faster beams having smaller durations.  To be consistent with this result, the FWHM of the radio emission must be related to the Langmuir waves generated by the fastest electrons.  The wave-wave processes that convert Langmuir waves into radio waves could be responsible.  Whilst simulating the radio waves from a propagating electron beam that had an initial Maxwellian distribution, \citet{Li:2009aa} found that an increased beam density resulted in a decreased duration of second harmonic emission.  Additionally, the duration of any observed type III emission will be affected by the scattering of radio waves from source to observer and this must be taken into account when composing a complete picture of electron beam transport.

The results from the simulations highlight why there is such a large standard deviation in the rise and decay times derived from the LOFAR type III bursts.  Electron beams with varying properties can produce radio bursts with significantly different characteristics.  Figure \ref{figure:free_streaming} suggests that the initial beam densities will play a significant role in determining the relevant velocity of electrons that drive the radio emission and might be one reason why the durations that we found with LOFAR are smaller than previous observational reports.  However, other different beam and background plasma parameters are also likely to influence the type III durations, including the beam spectral index, the background density profile and thermal velocity.

\section{Conclusions} \label{sec:conclusion}

We analysed the temporal and spectral profile of 31 isolated type III bursts that occurred between May--September 2015 observed with LOFAR.  The increased resolution allowed us to characterise the rise and decay times, together with the drift rates and the bandwidth in unprecedented detail.  We found that the rise and decay times associated with the type III bursts increase as a function of decreasing frequency.  Correspondingly the FWHM duration of the type III bursts increases as a function of decreasing frequency, in-line with previous observations, although the durations we found were slightly smaller than previously reported in the literature.

The drift rates of the type III bursts were simultaneously presented for the rise, peak and decay times.  Type III rise times resulted in higher magnitude drift rates than from decay times.  Drift rates were used to infer the velocities of the front, middle and back of the electron beams responsible for the radio emission.  The mean velocities found over all radio bursts observed are $0.20$~c, $0.17$~c, $0.15$~c for the front, middle and back of the beam, respectively.  This highlights the long-standing belief that the electron beam elongates in space due to a range of velocities within the electron beam.  What was particularly interesting was that the difference between the velocities at the front and the back of the beam, the rate of beam expansion in space, was proportional to the velocity of the middle of the beam. This can be interpreted that faster electron beams expand faster in space.  However, the presence of radio wave scattering and refraction strongly affects the observed type III burst characteristics.  Specifically, it could induce frequency dependent delay, which could lead to frequency dependent apparent drift \citep{Kontar:2017ab}.  The decay of the burst could be stronger affected by radio-wave propagation effects, reducing more the apparent speed at the back of the electron beam.

Despite moving with one characteristic velocity, electron beams at one point in space are not mono-energetic but are ensembles of electron at different energies.  We showed the difference between the limiting cases of free-streaming electrons and the gas-dynamic theory, where electrons propagate in a single beam-plasma structure.  We explained why neither theories are adequate to model the increasing duration of type III bursts as a function of decreasing frequency.  Using simulations, we demonstrated how the different velocities at the front and back of the electron beam are likely caused by ensembles of electrons with different minimum and maximum velocities; the elongation of the beam being caused by the collection of electrons at the front having higher energies than at the back.  

We showed that the duration inferred from the Langmuir waves is sensitive to the initial number density of the electron beam.  Variation in the initial electron beam and the background coronal parameters are likely to be significant in explaining the smaller durations that we found in comparison to previous observations.  However, the duration of type III bursts will also be affected by the wave-wave processes that convert Langmuir waves to radio waves and the scattering effects during radio wave propagation.  Our results increase the motivation to understand all the plasma physics that governs the temporal and spectral type III profiles, so that these radio bursts can be better used to remote sense electron beam properties in the solar corona and the inner heliosphere.

\begin{acknowledgements}
We acknowledge support from the STFC consolidated grant 173869-01.  Support from a Marie Curie International Research Staff Exchange Scheme Radiosun PEOPLE-2011-IRSES-295272 RadioSun project is greatly appreciated.  This work benefited from the Royal Society grant RG130642.  This paper is based (in part) on data obtained with the International LOFAR Telescope (ILT) under project code LC3-012 and LC4-016. LOFAR (van Haarlem et al. 2013) is the Low Frequency Array designed and constructed by ASTRON. It has observing, data processing, and data storage facilities in several countries, that are owned by various parties (each with their own funding sources), and that are collectively operated by the ILT foundation under a joint scientific policy. The ILT resources have benefitted from the following recent major funding sources: CNRS-INSU, Observatoire de Paris and Universit\'{e} d'Orléans, France; BMBF, MIWF-NRW, MPG, Germany; Science Foundation Ireland (SFI), Department of Business, Enterprise and Innovation (DBEI), Ireland; NWO, The Netherlands; The Science and Technology Facilities Council, UK.  We thank the staff of ASTRON, in particular Richard Fallows, for assistance in the set-up and testing for the solar LOFAR observations used.
\end{acknowledgements}

\bibliographystyle{aa}
\bibliography{/Users/hamish/Documents/papers/ubib}

\end{document}